# MANIN PAIRS AND TOPOLOGICAL FIELD THEORY

E. GETZLER

The Lie algebra of first order differential operators on the circle, sometimes known as the Atiyah algebra, has a central extension parametrized by a real number $d$, the chiral algebra spanned by the stress-energy tensor $\mathsf{T}(z)$ and current $\mathsf{J}(z)$, with singular operator products

$$\mathsf{T}(z) \cdot \mathsf{T}(w) \sim \frac{\frac{3}{2}d}{(z-w)^4} + \frac{2\mathsf{T}(w)}{(z-w)^2} + \frac{\partial \mathsf{T}(w)}{z-w},$$

$$\mathsf{T}(z) \cdot \mathsf{J}(w) \sim \frac{\mathsf{J}(w)}{(z-w)^2} + \frac{\partial \mathsf{J}(w)}{z-w},$$

$$\mathsf{J}(z) \cdot \mathsf{J}(w) \sim \frac{d}{(z-w)^2}.$$

We say that a field $A(z)$ has weight $(a, q)$ if it has conformal dimension $a$ and charge $q$:

$$\mathsf{T}(z) \cdot A(w) \sim \frac{aA(w)}{(z-w)^2} + \frac{\partial A(w)}{z-w},$$

$$\mathsf{J}(z) \cdot A(w) \sim \frac{qA(w)}{z-w}.$$

Kazama describes in [14] a chiral algebra spanned by the fields $\mathsf{T}(z)$ and $\mathsf{J}(z)$, together with fermionic fields $\mathsf{G}^+(z)$, $\mathsf{G}^-(z)$ and $\Phi(z)$ of weight $(\frac{3}{2}, 1)$, $(\frac{3}{2}, -1)$ and $(\frac{3}{2}, -3)$ and a bosonic field $\mathsf{F}(z)$ of weight $(2, -2)$. These fields have singular operator products

$$\mathsf{G}^+(z) \cdot \mathsf{G}^-(w) \sim \frac{d}{(z-w)^3} + \frac{\mathsf{J}(w)}{(z-w)^2} + \frac{\mathsf{T}(w) + \frac{1}{2}\partial \mathsf{J}(w)}{z-w},$$

$$\mathsf{G}^+(z) \cdot \Phi(w) \sim -\tfrac{1}{2}\mathsf{G}^-(z) \cdot \mathsf{G}^-(w) \sim \frac{F(w)}{z-w},$$

$$\mathsf{G}^+(z) \cdot \mathsf{G}^+(w) \sim \mathsf{G}^-(z) \cdot \Phi(w) \sim \mathsf{G}^+(z) \cdot F(w) \sim 0,$$

$$\mathsf{G}^-(z) \cdot F(w) \sim \frac{3\Phi(w)}{(z-w)^2} + \frac{\partial \Phi(w)}{z-w}.$$

With $\Phi(z)$ and $F(z)$ set to zero, this chiral algebra becomes the $N = 2$ superconformal algebra: for this reason, we will call it the $N = 1\frac{1}{2}$ superconformal algebra.

In this paper, we construct actions of the $N = 1\frac{1}{2}$ superconformal algebra, with as special cases the Kazama-Suzuki model, the $G/G$ model, and several deformations of

The author is partially supported by a fellowship of the Sloan Foundation, and a research grant of the NSF.

these models. (The $G/G$ model has also been considered by Isidro-Ramallo [11].) The data from which we construct such an action are as follows:

(1) a finite-dimensional reductive Lie algebra $\mathfrak{g}$ with invariant inner product $(-,-)$;
(2) a polarization $\mathfrak{g} = \mathfrak{g}_+ \oplus \mathfrak{g}_-$ of $\mathfrak{g}$ (that is, a decomposition into a direct sum of transverse isotropic subspaces) such that $\mathfrak{g}_+$ is a Lie subalgebra of $\mathfrak{g}$;
(3) a highest weight representation $\mathcal{E}$ of the affine Kac-Moody algebra $\hat{\mathfrak{g}}$ with central extension represented by the operator product expansion

$$X(z) \cdot Y(w) \sim \frac{(X,Y) - \frac{1}{2}\langle X,Y \rangle}{(z-w)^2} + \frac{[X,Y](w)}{z-w},$$

where $\langle -,- \rangle$ is the Killing form of $\mathfrak{g}$;
(4) an element $\alpha \in \mathfrak{g}_0$, where $\mathfrak{g}_0$ is the intersection of the orthogonal complements of $[\mathfrak{g}_+, \mathfrak{g}_+]$ and $[\mathfrak{g}_-, \mathfrak{g}_-]$.

The vector space underlying this representation of the $N = 1\frac{1}{2}$ algebra is the tensor product of the highest weight representation $\mathcal{E}$ and the fermionic Fock space associated to the polarized inner product space $\mathfrak{g}$. The central charge is

$$d = \tfrac{1}{2} \dim \mathfrak{g} - (\rho,\rho) - (\alpha,\alpha),$$

where $\rho = [x_i, x^i] \in \mathfrak{g}$; here $x_i$ and $x^i$ are dual bases of $\mathfrak{g}_+$ and $\mathfrak{g}^-$, and we employ the summation convention for repeated indices.

The data (1) and (2) are known as a Manin pair, and have been introduced by Drinfeld in the study of quasi-Hopf algebras [5]. If in addition $\mathfrak{g}_-$ is a Lie subalgebra of $\mathfrak{g}$, we say that $(\mathfrak{g}, \mathfrak{g}_+, \mathfrak{g}_-)$ form a Manin triple; this notion lies at the foundations of the Leningrad school's approach to completely integrable systems Semenov-Tian-Shansky [20]. In this case, we obtain an action of the $N = 2$ superconformal algebra. When $\alpha = 0$, this action has been constructed by Spindel, Sevrin, Troust and van Proyen ([22], [21]); the model of Kazama-Suzuki [15] is a special case of their results. Note that in the case of a Manin triple, the element $\alpha \in \mathfrak{g}_0$ corresponds to a pair of characters $x_\pm \mapsto (\alpha, x_\pm)$ of the two Lie algebras $\mathfrak{g}_\pm$.

In Section 7, we will construct an equivalence between the $\mathrm{SL}(2)/\mathrm{SL}(2)$ model and the $\mathrm{SL}(2)/\mathrm{SO}(1)$ model with a non-zero value of $\alpha$: the resulting equivalence of $N = 1\frac{1}{2}$ superconformal algebras generalizes the results of Aharony-Ganoor-et al. [1], Hu and Yu [10] and Frenkel [7], who showed that the cohomology of the two theories is isomorphic.

In an appendix, we explain the formulas which are used in the manipulation of operator products: while straightforward, their application leads to very complicated formulas, and the chances of making an error are greatly reduced by the availability of an excellent Mathematica package (Thielemans [23]). We also prove a result, found independently by Figueroa-O'Farill [6], which gives a small number of relations between fields $\mathsf{G}^\pm$ and $\Phi$ which must be verified in order for them to generate an action of the $N = 1\frac{1}{2}$ superconformal algebra.

This work was inspired by the article of Parkhomenko [17], who drew attention to the central role played in the work of Sevrin et al. [22] and [21] of Manin triples. We are also grateful to Bong Lian and Gregg Zuckerman for a number of interesting conversations.

1. THE $N = 1\frac{1}{2}$ SUPERCONFORMAL ALGEBRA

Recall the definition of the $N = 1$ superconformal algebra: this is the chiral algebra spanned by the stress-energy tensor $\mathsf{T}(z)$ and a fermionic field $\mathsf{G}(z)$, and their derivatives, with operator products

$$\mathsf{T}(z) \cdot \mathsf{T}(w) \sim \frac{\frac{3}{2}d}{(z-w)^4} + \frac{2\mathsf{T}(w)}{(z-w)^2} + \frac{\partial \mathsf{T}(w)}{z-w},$$

$$\mathsf{T}(z) \cdot \mathsf{G}(w) \sim \frac{\frac{3}{2}\mathsf{G}(w)}{(z-w)^2} + \frac{\partial \mathsf{G}(w)}{z-w},$$

$$\mathsf{G}(z) \cdot \mathsf{G}(w) \sim \frac{2d}{(z-w)^3} + \frac{2\mathsf{T}(w)}{z-w}.$$

Here, $d$ is a real number called the central charge of the algebra. (This normalization of the central charge, related to the usual central charge $c$ by the formula $d = c/3$, is more natural in the study of $N = 2$ models.) This chiral algebra may be embedded in the $N = 1\frac{1}{2}$ superconformal algebra, by mapping $\mathsf{G}$ to $\mathsf{G}^+ + \mathsf{G}^- + \Phi$ and $\mathsf{T}$ to $\mathsf{T}$.

Superfields allow us to express these operator products in a compact way. In the complex superspace with coordinates $Z = (z, \theta)$, introduce the differential operator

$$\mathcal{D} = \frac{\partial}{\partial \theta} + \theta \frac{\partial}{\partial z},$$

so that $\mathcal{D}^2 = \partial$. If $n \in \mathbb{Z}$, $\varepsilon \in \{0, 1\}$ and $W = (w, \xi)$, it is convenient to adopt the notations

$$\frac{\mathcal{A}(W)}{(Z-W)^n} = \frac{\mathcal{A}(W)}{(z-w-\theta\xi)^n} = \left((z-w)^{-n} + n\theta\xi(z-w)^{-n-1}\right)\mathcal{A}(W),$$

$$\frac{\mathcal{A}(W)}{(Z-W)^{n-\frac{1}{2}}} = \frac{2(\theta-\xi)\mathcal{A}(W)}{(z-w-\theta\xi)^n} = \frac{2(\theta-\xi)\mathcal{A}(W)}{(z-w)^n}.$$

Introducing the superfield $\mathcal{G}(Z) = \mathsf{G}(z) + 2\theta\mathsf{T}(z)$, the defining relations of the $N = 1$ superconformal algebra are contained in the singular operator product

$$\mathcal{G}(Z) \cdot \mathcal{G}(W) \sim \frac{2d}{(Z-W)^3} + \frac{\frac{3}{2}\mathcal{G}(W)}{(Z-W)^{3/2}} + \frac{\mathcal{D}\mathcal{G}(W)}{Z-W} + \frac{\partial\mathcal{G}(W)}{(Z-W)^{1/2}}.$$

A superfield $\Omega$ is a primary of dimension $\Delta$ if

$$\mathcal{G}(Z) \cdot \Omega(W) \sim \frac{\Delta\Omega(W)}{(Z-W)^{3/2}} + \frac{\mathcal{D}\Omega(W)}{Z-W} + \frac{\partial\Omega(W)}{(Z-W)^{1/2}}.$$

The $N=1$ superconformal algebra may be realized on the Fock space of a scalar field $\phi(z)$, with singular operator product

$$\phi(z) \cdot \phi(w) \sim -\log(z-w),$$

tensored with the Fock space of a free fermionic field $\psi(z)$ with singular operator product

$$\psi(z) \cdot \psi(w) \sim \frac{1}{z-w}.$$

The superfield $\Psi(Z) = \psi(z) + i\theta\partial\phi(z)$ is a primary of dimension $1/2$, and

$$\Psi(Z) \cdot \Psi(W) \sim \frac{1}{Z-W}.$$

The superfield $\mathsf{G} = \Psi\mathcal{D}\Psi$ realizes the $N=1$ superconformal algebra with central charge $d=1/2$, and the fields $\mathsf{T}$ and $\mathsf{G}$ are given by the formulas

$$\mathsf{T} = -\tfrac{1}{2}(\partial\phi)^2 - \tfrac{1}{2}\psi\,\partial\psi,$$
$$\mathsf{G} = i\partial\phi\,\psi.$$

The following simple result is an analogue for the $N=1$ superconformal algebra of the well-known deformation $\mathsf{T} + \alpha\partial J$ of a stress-energy tensor by the derivative of a current.

**Proposition 1.1.** *If $\mathsf{G}$ is a superfield realizing the $N=1$ superconformal algebra with central charge $d$, and $\Phi$ is a primary fermionic superfield of dimension $1/2$ such that*

$$\Psi(Z) \cdot \Psi(W) \sim \frac{1}{Z-W},$$

*then $\mathsf{G} + \alpha\partial\Psi$ realizes the $N=1$ superconformal algebra with central charge $d-\alpha^2$.*

In the case of the free field realization, we obtain a deformation with central charge $d-\alpha^2$, given explicitly by the formulas

$$\mathsf{T} = -\tfrac{1}{2}(\partial\phi)^2 - \tfrac{1}{2}\psi\,\partial\psi + \tfrac{i}{2}\alpha\partial^2\phi,$$
$$\mathsf{G} = i\partial\phi\,\psi + \alpha\partial\psi.$$

The free-field realization of the $N=1$ superconformal algebra has a generalization, in which the place of the vector space $\mathbb{R}$ is taken by a reductive Lie algebra $\mathfrak{g}$ with invariant inner product $(-,-)$. Let $x_i$ be a basis of $\mathfrak{g}$, relative to which $[x_i, x_j] = C_{ij}^k$ and $h_{ij} = (x_i, x_j)$, with matrix inverse $h^{ij}$. Denote $h_{kl}C_{ij}^l$ by $C_{ijk}$: it is antisymmetric in all three indices, reflecting the invariance of the inner product.

Let $\langle -,- \rangle$ be the Killing form of $\mathfrak{g}$, with coefficients

$$g_{ij} = \langle x_i, x_j \rangle = C_{ik}^l C_{jl}^k.$$

Let $\mathfrak{g}$ be a simple Lie algebra with Cartan subalgebra $\mathfrak{h}^*$ and highest root $\theta$. Let $h$ be the dual Coxeter number of $\mathbf{k}$. We say that an inner product $(-,-)$ has level $k$ if

$$\frac{(x,x)}{k+h} = \frac{\langle x,x \rangle}{2h}.$$

The Killing form itself has level $h$.

Consider the chiral algebra generated by fields $J_i(z)$ satisfying the operator product expansions

$$J_i(z) \cdot J_j(w) = \frac{h_{ij} - \frac{1}{2}g_{ij}}{(z-w)^2} + \frac{C_{ij}^k J_k(w)}{z-w}.$$

These relations define a representation of the affine Kac-Moody algebra $\widehat{\mathfrak{g}}$, with central extension corresponding to the invariant inner product $(x,y) - \frac{1}{2}\langle x,y \rangle$ on $\mathfrak{g}$. Introduce free fermionic fields $a^i(z)$, with operator products

$$a^i(z) \cdot a^j(w) \sim \frac{h^{ij}}{z-w}.$$

The superfields $\mathcal{J}_i(Z) = h_{ij}a^j(z) + \theta(J_i - \frac{1}{2}C_{ijk}a^j a^k)$ realize the super Kac-Moody algebra associated to $\mathfrak{g}$ (Kac-Todorov [12]),

$$\mathcal{J}_i(Z) \cdot \mathcal{J}_j(W) \sim \frac{h_{ij}}{Z-W} + \frac{\frac{1}{2}C_{ij}^k \mathcal{J}_k(w)}{(Z-W)^{1/2}}.$$

The fields

$$\mathsf{G} = J_i a^i - \tfrac{1}{6} C_{ijk} a^i a^j a^k,$$
$$\mathsf{T} = \tfrac{1}{2}\left(h^{ij} J_i J_j + h_{ij} \partial a^i a^j\right),$$

realize the $N=1$ superconformal algebra, with central charge $d = \frac{1}{2}\dim \mathfrak{g} - \frac{1}{6}h_{ij}g^{ij}$. When $\mathfrak{g}$ is abelian, the Killing form vanishes, and this model becomes the free-field realization of the $N=1$ superconformal chiral algebra.

The superfield $\mathcal{J}$ is a primary of dimension 1; it follows that if $\alpha$ is an element of $\mathfrak{g}$, we may deform $\mathcal{G}$ to $\mathcal{G} + (\alpha, \mathcal{DJ})$, obtaining a realization of the $N=1$ superconformal algebra with central charge $d = \frac{1}{2}\dim \mathfrak{g} - \frac{1}{6}h_{ij}g^{ij} - (\alpha,\alpha)$.

The problem which we discuss in this paper is the construction of actions of the $N = 1\frac{1}{2}$ superconformal symmetry underlying this $N=1$ superconformal symmetry.

We may assemble the fields of the $N = 1\frac{1}{2}$ superconformal algebra into superfields by the following formulas:

$$\mathcal{G}(Z) = (\mathsf{G}^+(z) + \mathsf{G}^-(z) + \Phi(z)) + 2\theta \mathsf{T}(z),$$
$$\mathcal{J}(Z) = -\mathsf{J}(z) + \theta(\mathsf{G}^+(z) - \mathsf{G}^-(z) - 3\Phi(z)),$$
$$\mathcal{F}(Z) = \Phi(z) + \theta \mathsf{F}(z).$$

Then $\mathcal{G}$ generates an $N = 1$ superconformal symmetry, $\mathcal{J}$ and $\mathcal{F}$ are bosonic and fermionic superfields of dimension 1 and $\frac{3}{2}$, and

$$\mathcal{J}(Z) \cdot \mathcal{J}(W) \sim \frac{d}{(Z-W)^2} + \frac{\frac{1}{2}\mathcal{G}(W) + \mathcal{F}(W)}{(Z-W)^{1/2}},$$

$$\mathcal{J}(Z) \cdot \mathcal{F}(W) \sim \frac{3\mathcal{F}(W)}{Z-W} + \frac{\frac{1}{2}D\mathcal{F}(W)}{(Z-W)^{1/2}},$$

$$\mathcal{F}(Z) \cdot \mathcal{F}(W) \sim 0.$$

We may think of $\mathcal{F}$ as a superfield representing a curvature term.

When the curvature $\mathcal{F}$ equals zero and we have the $N = 2$ superconformal algebra, there is an automorphism which sends $\mathcal{J}$ to $-\mathcal{J}$, and leaves $\mathcal{G}$ fixed. In terms of the component fields, this exchanges $\mathsf{G}^+$ and $\mathsf{G}^-$, and sends $\mathsf{J}$ to $-\mathsf{J}$. There is no analogue of this automorphism when $\mathcal{F}$ is non-zero.

The following result, which has been found independently by Figueroa-O'Farill [6] when $\Phi = \mathsf{F} = 0$, simplifies the verification of these operator products in concrete situations. The proof is contained in the Appendix, along with some background on vertex algebras used in its proof.

**Proposition 1.2.** *Let $\mathsf{G}^\pm(z)$ and $\Phi(z)$ be fermionic fields such that*

$$\mathsf{G}^+(z) \cdot \mathsf{G}^-(w) \sim \frac{d}{(z-w)^3} + \frac{\mathsf{J}(w)}{(z-w)^2} + \frac{\mathsf{T}(w) + \frac{1}{2}\partial\mathsf{J}(w)}{z-w},$$

$$\mathsf{G}^+(z) \cdot \Phi(w) \sim -\tfrac{1}{2}\mathsf{G}^-(z) \cdot \mathsf{G}^-(w) \sim \frac{\mathsf{F}(w)}{z-w},$$

$$\mathsf{G}^+(z) \cdot \mathsf{G}^+(w) \sim 0,$$

*for a scalar $d$ and bosonic fields $\mathsf{J}(z)$, $\mathsf{T}(z)$ and $\mathsf{F}(z)$, and*

$$\mathsf{J}(z) \cdot \mathsf{G}^\pm(w) \sim \pm\frac{\mathsf{G}^\pm(w)}{z-w}, \quad \mathsf{J}(z) \cdot \Phi(w) \sim -\frac{3\Phi(w)}{z-w}.$$

*Then these fields satisfy the operator products of the $N = 1\frac{1}{2}$ superconformal algebra, with central charge $d$.*

The following corollary applies to the case when the curvature $\mathcal{F}$ vanishes.

**Corollary 1.3.** *Let $\mathsf{G}^\pm(z)$ be fermionic fields such that $\mathsf{G}^\pm(z) \cdot \mathsf{G}^\pm(w) \sim 0$,*

$$\mathsf{G}^+(z) \cdot \mathsf{G}^-(w) \sim \frac{d}{(z-w)^3} + \frac{\mathsf{J}(w)}{(z-w)^2} + \frac{\mathsf{T}(w) + \frac{1}{2}\partial\mathsf{J}(w)}{z-w},$$

*for a scalar $d$ and bosonic fields $\mathsf{J}(z)$ and $\mathsf{T}(z)$, and*

$$\mathsf{J}(z) \cdot \mathsf{G}^\pm(w) \sim \pm\frac{\mathsf{G}^\pm(w)}{z-w}.$$

*Then these fields satisfy the operator products of the $N = 2$ superconformal algebra.*

The following result provides an analogue of Proposition 1.1 for the $N = 1\tfrac{1}{2}$ superconformal algebra.

**Proposition 1.4.** *Suppose $\Omega^\pm(Z)$ are fermionic superfields such that*

$$\mathcal{G}(Z) \cdot \Omega^\pm(W) \sim \frac{\tfrac{1}{2}\Omega^\pm(W)}{(Z-W)^{3/2}} + \frac{\mathcal{D}\Omega^\pm(W)}{Z-W} + \frac{\partial\Omega^\pm(W)}{(Z-W)^{1/2}},$$

$$\mp \mathcal{J}(Z) \cdot \Omega^\pm(W) \sim \frac{\Omega^\pm(W)}{Z-W} + \frac{\tfrac{1}{2}\mathcal{D}\Omega^\pm(W)}{(Z-W)^{1/2}},$$

$$\Omega^\pm(Z) \cdot \Omega^\pm(W) \sim \mathcal{F}(Z) \cdot \Omega^\pm(W) \sim 0,$$

$$\Omega^+(Z) \cdot \Omega^-(W) \sim \frac{d'}{Z-W}.$$

*Then $\widetilde{\mathcal{G}} = \mathcal{G} + \partial\Omega^+ + \partial\Omega^-$, $\widetilde{\mathcal{J}} = \mathcal{J} - \mathcal{D}\Omega^+ + \mathcal{D}\Omega^-$ and $\widetilde{\mathcal{F}} = \mathcal{F}$ define a realization of the $N = 1\tfrac{1}{2}$ superconformal algebra with central extension $d - 2d'$.*

*Proof.* Write $\Omega^\pm(Z) = A^\pm(z) + \theta B^\pm(z)$. Then

$$\mathsf{G}^\pm(z) \cdot A^\pm(w) \sim \frac{B^\pm(w)}{z-w},$$

$$\mathsf{G}^\mp(z) \cdot A^\pm(w) \sim \mathsf{G}^\pm(z) \cdot B^\pm(w) \sim 0,$$

$$\mathsf{G}^\mp(z) \cdot B^\pm(w) \sim \frac{A^\pm(w)}{(z-w)^2} + \frac{\partial A^\pm(w)}{z-w},$$

$$A^+(z) \cdot A^-(w) \sim \frac{d'}{z-w}, \quad B^+(z) \cdot B^-(w) \sim \frac{d'}{(z-w)^2},$$

while $\Phi(z) \cdot A^\pm(w) \sim A^\pm(z) \cdot B^\pm(w) \sim A^\pm(z) \cdot B^\mp(w) \sim 0$. The deformed fields are given by the formulas

$$\widetilde{\mathsf{T}} = \mathsf{T} + \tfrac{1}{2}(\partial B^+ + \partial B^-), \qquad \widetilde{\mathsf{G}}^\pm = \mathsf{G}^\pm + \partial A^\pm, \qquad \widetilde{\mathsf{J}} = \mathsf{J} + B^+ - B^-,$$

while $\widetilde{\Phi} = \Phi$ and $\widetilde{\mathsf{F}} = \mathsf{F}$. The verification of the hypotheses of Proposition 1.2 is now straightforward. $\square$

## 2. Manin pairs

The following definition is due to Drinfeld [5].

**Definition 2.1.** *Let $\mathfrak{g}$ be a finite dimensional, reductive Lie algebra over $\mathbb{C}$, with invariant inner product $(-,-)$. A Manin pair $(\mathfrak{g}, \mathfrak{g}_+, \mathfrak{g}_-)$ is a polarization $\mathfrak{g} = \mathfrak{g}_+ \oplus \mathfrak{g}_-$ of $\mathfrak{g}$ such that $\mathfrak{g}_+$ is a Lie subalgebra of $\mathfrak{g}$; it is a Manin triple if in addition $\mathfrak{g}_-$ is a Lie subalgebra of $\mathfrak{g}$.*

In fact, Drinfeld does not include the complementary isotropic subspace $\mathfrak{g}_-$ in the data defining a Manin pair: however, we find our more precise notion of a Manin pair to be convenient for the purposes of this paper.

Given a Manin pair, define the subspace $\mathfrak{g}_0 \subset \mathfrak{g}$ to be the intersection of the subspaces $[\mathfrak{g}_+, \mathfrak{g}_+]^\perp$ and $[\mathfrak{g}_-, \mathfrak{g}_-]^\perp$.

An example of a Manin pair is obtained by taking a reductive Lie algebra $\mathbf{k}$ with an invariant inner product $(-,-)$: we let $\mathfrak{g} = \mathbf{k} \oplus \mathbf{k}$, with inner product

$$(x_1 \oplus y_1, x_2 \oplus y_2) = (x_1, x_2) - (y_1, y_2).$$

Let $\mathfrak{g}_\pm = \{x \oplus \pm x \mid x \in \mathbf{k}\}$: it is clear that these subspaces are isotropic, and that $\mathfrak{g}_+$ is a Lie subalgebra (though not $\mathfrak{g}_-$, since $[\mathfrak{g}_-, \mathfrak{g}_-] \subset \mathfrak{g}_+$). This Manin pair is associated, by the construction of this paper, to the action of the $N = 1\frac{1}{2}$ superconformal algebra in the $G/G$ model.

Examples of Manin triples may be obtained by taking the following data:

(1) a simple Lie algebra $\mathbf{k}$, with Borel subalgebra $\mathbf{b}$;
(2) an invariant inner product $(-,-)$ on $\mathbf{k}$;
(3) a parabolic subalgebra $\mathbf{p} \supset \mathbf{b}$.

The parabolic subalgebra $\mathbf{p}$ has the Levi decomposition $\mathbf{p} = \mathbf{l} \oplus \mathbf{n}$, where $\mathbf{l}$ is reductive and $\mathbf{n}$ is nilpotent. Let $\mathfrak{g}$ be the Lie algebra $\mathfrak{g} = \mathbf{k} \oplus \mathbf{l}$, with inner product

$$(x_1 \oplus y_1, x_2 \oplus y_2) = (x_1, x_2) - (y_1, y_2).$$

The subalgebra

$$\mathfrak{g}_+ = \{x \oplus y \in \mathbf{k} \oplus \mathbf{l} \mid x - y \in \mathbf{n}\} \subset \mathfrak{g}$$

is easily seen to be isotropic, and $\dim \mathfrak{g}_+ = \frac{1}{2} \dim \mathfrak{g}$.

Let $\mathfrak{h}$ be the Cartan subalgebra of $\mathbf{k}$ determined by the Borel subalgebra $\mathbf{b}$; if $x \in \mathbf{k}$, denote by $x_\mathfrak{h}$ its projection onto $\mathfrak{h}$. Let $\mathbf{b}_-$ be the Borel subalgebra opposite to $\mathbf{b}$. The subalgebra

$$\mathfrak{g}_- = \{x \oplus y \in (\mathbf{k} \cap \mathbf{b}_-) \oplus (\mathbf{l} \cap \mathbf{b}) \mid (x+y)_\mathfrak{h} = 0\} \subset \mathfrak{g}$$

is isotropic, and together with $\mathfrak{g}_+$ forms a Manin triple.

The two extreme cases of this construction are of especial interest:

(1) If $\mathbf{p} = \mathbf{b}$ is itself a Borel subalgebra, then $\mathfrak{g} = \mathbf{k} \oplus \mathfrak{h}$ and

$$\mathfrak{g}_\pm = \{x \oplus h \in \mathbf{b}_\pm \oplus \mathfrak{h} \mid x_\mathfrak{h} = \pm h\}.$$

This Manin triple corresponds to the Kazama-Suzuki model.

(2) If $\mathbf{p} = \mathbf{k}$ is all of $\mathbf{k}$, then $\mathfrak{g} = \mathbf{k} \oplus \mathbf{k}$ and

$$\mathfrak{g}_+ = \{x \oplus x \in \mathbf{k} \oplus \mathbf{k}\},$$
$$\mathfrak{g}_- = \{x \oplus y \in \mathbf{b}_- \oplus \mathbf{b}_+ \mid x_\mathfrak{h} + y_\mathfrak{h} = 0\}.$$

This Manin triple corresponds to the $G/G$ model (and is also familiar as the Manin triple associated to the classical limit of the quantum group $U_q\mathbf{k}$).

Denote by $\Delta \subset \mathfrak{h}^*$ the roots of $\mathbf{k}$, and by $\tilde{\Delta} \subset \Delta$ the roots of $\mathbf{l} \subset \mathbf{p}$. Let $\Pi \subset \Delta$ be the basis of the root system $\Delta$ determined by $\mathbf{b}$, and let $\Delta = \Delta_+ \cup \Delta_-$ be the associated decomposition of $\Delta$ into positive and negative roots. Then $\mathbf{p}$ has the decomposition into a sum of weight spaces of $\mathbf{k}$,

$$\mathbf{p} = \sum_{\alpha \in \Delta_+ \cup \tilde{\Delta}} \mathbf{k}_\alpha,$$

where for $\alpha \in \Delta$, $\mathbf{k}_\alpha = \{x \in \mathbf{k} \mid Hx = \alpha(H)x \text{ for all } H \in \mathfrak{h}\}$.

To a root $\alpha \in \Delta \subset \mathfrak{h}^*$ is associated a coroot $\alpha\check{} \in \Delta\check{} \subset \mathfrak{h}$, such that the reflection $s_\alpha$ is given by the formula

$$s_\alpha \beta = \beta - \alpha(\beta\check{});$$

in particular, $\alpha(\alpha\check{}) = 2$. For $\alpha \in \Delta_+$, let $x_\alpha \in \mathbf{k}_\alpha$ and $x_{-\alpha} \in \mathbf{k}_{-\alpha}$ be vectors such that

$$[x_\alpha, x_{-\alpha}] = \alpha\check{}, \quad [\alpha\check{}, x_{\pm\alpha}] = \pm 2 x_\alpha.$$

It follows that

$$(x_\alpha, x_{-\alpha}) = \tfrac{1}{2}([\alpha\check{}, x_\alpha], x_{-\alpha}) = \tfrac{1}{2}(\alpha\check{}, [x_\alpha, x_{-\alpha}]) = \tfrac{1}{2}(\alpha\check{}, \alpha\check{}).$$

Given the Manin triple $(\mathfrak{g} = \mathbf{k} \oplus \mathbf{l}, \mathfrak{g}_+, \mathfrak{g}_-)$, $\mathfrak{g}_+$ has basis

$$\left\{ x_\alpha \oplus 0 \mid \alpha \in \Delta_+ \backslash \tilde{\Delta}_+ \right\} \cup \left\{ \alpha_i\check{} \oplus \alpha_i\check{} \mid \alpha_i \in \Pi \right\} \cup \left\{ x_{\pm\alpha} \oplus x_{\pm\alpha} \mid \alpha \in \tilde{\Delta}_+ \right\}.$$

If $\{\alpha_i \mid 1 \leq i \leq r\}$ are the simple roots of $\mathbf{k}$, the fundamental weights $\{\omega_i \mid 1 \leq i \leq r\} \subset \mathfrak{h}^*$ are the dual basis of the basis $\{\alpha\check{}_i \mid 1 \leq i \leq r\}$ of $\mathfrak{h}$. We may identify the weights $\omega_i$ with elements of $\mathfrak{h}$ by means of the restriction to $\mathfrak{h}$ of the inner product $(-,-)$. With these notations, the basis of $\mathfrak{g}_-$ dual to the above basis of $\mathfrak{g}_+$ is

$$\left\{ \tfrac{2}{(\alpha\check{},\alpha\check{})}(x_{-\alpha} \oplus 0) \mid \alpha \in \Delta_+ \right\} \cup \left\{ \tfrac{1}{2}(\omega_i \oplus -\omega_i) \mid 1 \leq i \leq r \right\} \cup \left\{ \tfrac{2}{(\alpha\check{},\alpha\check{})}(0 \oplus -x_\alpha) \mid \alpha \in \tilde{\Delta}_+ \right\}.$$

Thus, $\mathfrak{g}_0$ has the basis

$$\left\{ \alpha_i\check{} \oplus \alpha_i\check{} \mid \alpha_i \in \Pi \right\} \cup \left\{ \omega_i \oplus -\omega_i \mid \alpha_i \in \Pi \cap \tilde{\Delta} \right\}.$$

If $(\mathfrak{g}, \mathfrak{g}_+, \mathfrak{g}_-)$ is a Manin triple, the inner product on $\mathfrak{g}$ induces an identification of $\mathfrak{g}_-$ with the dual $\mathfrak{g}_+^*$ of $\mathfrak{g}_+$, and vice versa. The choice of a basis $x_i$ of $\mathfrak{g}_+$ determines a dual basis $x^i$ of $\mathfrak{g}_-$, such that the bracket of $\mathfrak{g}$ is given by the formulas

$$[x_i, x_j] = c_{ij}^k x_k,$$
$$[x_i, x^j] = f_i^{jk} x_k + c_{ki}^j x^k,$$
$$[x^i, x^j] = f_k^{ij} x^k - \phi^{ijk} x_k.$$

The coefficients $c_{ij}^k$, $f_k^{ij}$ and $\phi^{ijk}$ satisfy the identities

$$c_{ij}^k = -c_{ji}^k, \quad f_k^{ij} = -f_k^{ji}, \quad \phi^{ijk} = -\phi^{jik} = -\phi^{ikj},$$

$$c_{ij}^m c_{mk}^l + c_{jk}^m c_{mi}^l + c_{ki}^m c_{mj}^l = 0,$$

$$f_m^{ij} f_l^{mk} + f_m^{jk} f_l^{mi} + f_m^{ki} f_l^{mj} = c_{lm}^i \phi^{jkm} + c_{lm}^j \phi^{kim} + c_{lm}^k \phi^{ijm},$$

$$c_{mk}^i f_l^{jm} - c_{ml}^i f_k^{jm} - c_{mk}^j f_l^{im} + c_{ml}^j f_k^{im} = c_{kl}^m f_m^{ij},$$

$$f_m^{kl} \phi^{ijm} + f_m^{il} \phi^{jkm} + f_m^{jl} \phi^{kim} = 0.$$

Being a Manin triple is equivalent to the vanishing of the coefficients $\phi^{ijk}$.

Let us illustrate these formulas in the case of the Manin pair associated to $\mathbf{k} \oplus \mathbf{k}$. If we choose a basis $x_i$ of $\mathbf{k}$ with respect to which the inner product is given by the matrix $h_{ij}$, we obtain dual bases $x_i \oplus x_i$ and $\frac{1}{2} h^{ij}(x_j \oplus -x_j)$ of $\mathfrak{g}_+$ and $\mathfrak{g}_-$. The structure coefficients $c_{ij}^k$ are those of the Lie algebra $\mathbf{k}$, the coefficients $f_k^{ij}$ vanish, and the coefficients $\phi^{ijk}$ are given by the formula $\phi^{ijk} = \frac{1}{4} h^{il} h^{jm} c_{lm}^k$.

Given $x \in \mathfrak{g}$, denote its projection onto $\mathfrak{g}_\pm$ by $x_\pm$, and write $\bar{x} = x_+ - x_-$. Writing $x_+ = p^i x_i$ and $x_- = q_i x^i$, the condition that $x \in \mathfrak{g}_0$ is expressed by the formulas

$$p^i f_i^{jk} = q_i c_{jk}^i = q_i \phi^{ijk} = 0.$$

The element $\rho = [x_i, x^i] \in \mathfrak{g}$ is given by the explicit formula

$$\rho = f_j^{ji} x_i + c_{ij}^j x^i.$$

It is easily seen that $\rho \in \mathfrak{g}_0$.

Introduce the notations $\mathrm{ad}(x_i) x_j = c_{ij}^k x_k$ and $\mathrm{ad}^*(x^i) x_j = f_j^{ki} x_k$ for the adjoint action of $\mathfrak{g}_+$ on itself, and for the "action" of $\mathfrak{g}_-$ on $\mathfrak{g}_+$; $\mathrm{ad}(x_+)$ is a derivation of the Lie algebra $\mathfrak{g}_+$, while $\mathrm{ad}^*(x_-)$ is provided $x_-$ is orthogonal to $[\mathfrak{g}_+, \mathfrak{g}_+]$, in particular, if $x_- \in \mathfrak{g}_0$.

Let $D : \mathfrak{g}_+ \to \mathfrak{g}_+$ denote the derivation $D = -\mathrm{ad}(\rho_+) - \mathrm{ad}^*(\rho_-)$ of the Lie algebra $\mathfrak{g}_+$, given by the explicit formula

$$D x_i = (f_k^{jk} c_{ji}^l + c_{jk}^k f_i^{jl}) x_l.$$

**Lemma 2.2.** $D x_i = f_i^{kl} c_{kl}^j x_j$ and $\mathrm{Tr}(D) = -(\rho, \rho) = -2(\rho_+, \rho_-)$

*Proof.* Let $A : \mathfrak{g}_+ \otimes \mathfrak{g}_- \to \mathbb{R}$ be the bilinear form $A(x, y) = -\mathrm{Tr}(\mathrm{ad}(x) \mathrm{ad}^*(y))$, with matrix $A_i^j = c_{il}^k f_k^{jl}$. Summing over the indices $i$ and $l$ in the formula

$$c_{mk}^i f_l^{jm} - c_{ml}^i f_k^{jm} - c_{mk}^j f_l^{im} + c_{ml}^j f_k^{im} = c_{kl}^m f_m^{ij},$$

we see that

$$-A_k^j - c_{mi}^i f_k^{jm} - c_{mk}^j f_i^{im} - c_{lm}^j f_k^{lm} + A_k^j = 0.$$

The first and last terms cancel, and we obtain the formula for $Dx_i$. Taking the trace over the indices $j$ and $k$, we obtain the formula for $\mathrm{Tr}(D)$. □

Denote by $\langle -, - \rangle$ the Killing form of $\mathfrak{g}$, and by $\langle -, - \rangle_+$ the Killing form of $\mathfrak{g}_+$. The proofs of the following formulas are straightforward:

$$\langle x_i, x_j \rangle = 2\langle x_i, x_j \rangle_+, \quad \langle x_i, x^j \rangle = -D_i^j - 2A_i^j, \quad \langle x^i, x^j \rangle = 2f_l^{ik} f_k^{jl} + c_{kl}^i \phi^{jkl} + c_{kl}^j \phi^{ikl}.$$

We see that $\langle x_i, x^i \rangle = -3\operatorname{Tr}(D)$ and $\operatorname{Tr}(D^2) = \frac{1}{2}\langle \rho, \rho \rangle$.

Consider the Manin triple $(\mathfrak{g} = \mathbf{k} \oplus \mathbf{l}, \mathfrak{g}_+, \mathfrak{g}_-)$. Identify the weights

$$\rho(\mathbf{k}) = \frac{1}{2} \sum_{\alpha \in \Delta_+} \alpha \quad \text{and} \quad \rho(\mathbf{l}) = \frac{1}{2} \sum_{\alpha \in \tilde{\Delta}_+} \alpha$$

with elements of $\mathfrak{h}$ using the restriction of the inner product $(-, -)$ to $\mathfrak{h}$. From the explicit basis introduced above for $\mathfrak{g} = \mathbf{k} \oplus \mathbf{l}$, it follows that

$$\rho = \sum_{\alpha \in \Delta_+} \frac{2\alpha^{\smile}}{(\alpha^{\smile}, \alpha^{\smile})} \oplus \sum_{\alpha \in \tilde{\Delta}_+} \frac{2\alpha^{\smile}}{(\alpha^{\smile}, \alpha^{\smile})} = 2\rho(\mathbf{k}) \oplus 2\rho(\mathbf{l}),$$

so that $\rho_\pm = (\rho(\mathbf{k}) \pm \rho(\mathbf{l})) \oplus (\pm\rho(\mathbf{k}) + \rho(\mathbf{l}))$, and

$$\operatorname{Tr}(D) = -2(\rho(\mathbf{k}), \rho(\mathbf{k})) + 2(\rho(\mathbf{l}), \rho(\mathbf{l})).$$

In the case that $\mathbf{p} = \mathbf{b}$, Freudenthal's formula

$$\frac{(\rho, \rho)}{(\theta, \theta)} = \frac{h \dim \mathbf{k}}{24},$$

where $\theta$ is the highest root of $\mathbf{k}$, shows that

$$\operatorname{Tr}(D) = -\frac{h \dim \mathbf{k}}{6(k + h)}.$$

Finally, we recall Drinfeld's twisting operation [5]. Given a Manin pair $(\mathfrak{g}, \mathfrak{g}_+, \mathfrak{g}_-)$ and a linear map $R: \mathfrak{g}_i \to \mathfrak{g}_+$, we define a subspace

$$R \cdot \mathfrak{g}_- = \{x - Rx \mid x \in \mathfrak{g}_-\}.$$

Introducing the matrix elements $Rx^i = R^{ij}x_j$, the condition that $R \cdot \mathfrak{g}_-$ is isotropic is equivalent to the matrix $R^{ij}$ being antisymmetric. Taking as our dual bases of $\mathfrak{g}_+$ and $R \cdot \mathfrak{g}_-$ the vectors $x_i$ and $x^i - R^{ij}x_j$, we see that the structure coefficients of the Manin pair are modified in the following way:

$$c_{ij}^k \to c_{ij}^k,$$
$$f_k^{ij} \to f_k^{ij} - R^{il}c_{kl}^j - R^{lj}c_{kl}^i,$$
$$\phi^{ijk} \to \phi^{ijk} + R^{il}f_l^{jk} - R^{jl}f_l^{ik} - R^{il}R^{jm}c_{lm}^k.$$

We may illustrate this construction with the Manin triple with $\mathfrak{g} = \mathbf{k} \oplus \mathbf{k}$. Then $\mathfrak{g}_+$ has basis

$$\{X_{\pm\alpha} = x_{\pm\alpha} \oplus x_{\pm\alpha} \mid \alpha \in \Delta_+\} \cup \{X_i = \alpha^{\smile} \oplus \alpha^{\smile} \mid \alpha \in \Pi\},$$

while $\mathfrak{g}_-$ has the dual basis

$$\left\{X^{\pm\alpha} = \pm\frac{2x_{\mp\alpha}}{(\alpha\check{},\alpha\check{})} \mid \alpha \in \Delta_+\right\} \cup \left\{X^i = \tfrac{1}{2}(\omega_i \oplus -\omega_i) \mid 1 \leq i \leq r\right\}.$$

The twist

$$R = \sum_{\alpha \in \Delta_+} \frac{X_\alpha \wedge X_{-\alpha}}{(\alpha\check{},\alpha\check{})}$$

transforms this Manin triple to the Manin pair for which $\mathfrak{g}_- = \{(x,-x) \mid x \in \mathbf{k}\}$.

## 3. Finite-dimensional topological field theory

In this section, we describe a finite-dimensional analogue of the construction of the next section; this section may be skipped, since its results are not used elsewhere in this paper.

Let $A$ be a graded commutative algebra. A zeroth order differential operator is an operator which commutes with elements of $A$, while an $n$-th order differential operator on $A$ is an operator $L$ whose commutator with elements of $A$ is an $(n-1)$-th order differential operator. A first order derivation is a derivation in the usual sense, while an $n$-th order derivation on $A$ is an operator $L$ such that $[L,a] - La$ is an $(n-1)$-th order derivation for all $a \in A$ (see Penkava-Schwartz [18]). In particular, $n$-th order derivations are $n$-th order differential operators. A second-order derivation is an operator satisfying the formula

$$\Delta(abc) = \Delta(ab)c + (-1)^{|\Delta||a|}a\Delta(bc) + (-1)^{(|a|+|\Delta|)|b|}b\Delta(ac)$$
$$- (\Delta a)bc - (-1)^{|\Delta||a|}a(\Delta b)c - (-1)^{|\Delta|(|a|+|b|)}ab(\Delta c).$$

**Definition 3.1.** A Batalin-Vilkovisky algebra is a graded commutative algebra $A$ together with a second-order derivation $\Delta$ of degree $-1$ such that $\Delta^2 = 0$.

A differential Batalin-Vilkovisky algebra $(A,\Delta,\delta)$ is a Batalin-Vilkovisky algebra together with a derivation $\delta$ of degree 1 on $A$ such that $\delta^2 = 0$, and such that $\delta\Delta + \Delta\delta$ is a first-order differential operator.

Let $A$ be the exterior algebra $\Lambda \mathbf{k}^*$, where $\mathbf{k}$ is a vector space. Given a basis $x_i$ of $\mathbf{k}$ with dual basis $x^i$ of $\mathbf{k}^*$, we denote by $a_i$ the operator of degree $-1$ consisting of contraction with $x_i$, and by $a^i$ the operation of exterior multiplication by $x^i$. These operators satisfy the canonical commutation relations

$$a^i a_j + a_j a^i = \delta^i_j.$$

An operator $L$ on $A$ is an $n$-th differential operator if it can be written as a sum of monomials each of which is at most $n$-th order in the operators $\{a_i\}$.

A second order derivation $\Delta$ of degree $-1$ on $A$ has the form

$$\Delta = \tfrac{1}{2}f^{ij}_k a_i a_j a^k + p^i a_i$$

for coefficients $f_k^{ij}$ and $p^i$ such that $f_k^{ij} + f_k^{ji} = 0$. The author learnt the following lemma from G. Zuckerman.

**Lemma 3.2.** *The formula $\Delta^2 = 0$ holds if and only if $f_k^{ij}$ are the structure coefficients of a Lie coalgebra, $f_m^{ij} f_l^{mk} + f_m^{jk} f_l^{mi} + f_m^{ki} f_l^{mj} = 0$, and $p^k f_k^{ij} = 0$.*

*Proof.* We calculate

$$\Delta^2 = \tfrac{1}{4} f_k^{ij} f_{k'}^{i'j'} a_i a_j \big( \delta_{i'}^k a_{j'} - \delta_{j'}^k a_{i'} + a_{i'} a_{j'} a^k \big) a^{k'} + p^k f_k^{ij} a_i a_j.$$

The term proportional to $a^k a^{k'}$ vanishes, so that

$$\Delta^2 = \tfrac{1}{2} f_l^{mi} f_m^{jk} a^l a_i a_j a_k + p^k f_k^{ij} a_i a_j.$$

The first term is antisymmetric in $a_i$, $a_j$ and $a_k$, and we see that its vanishing is precisely the Jacobi rule for the Lie cobracket $x_i \mapsto \tfrac{1}{2} f_k^{ij} x_i \wedge x_j$ on $\mathbf{k}$. $\square$

A derivation of degree 1 of $A$ has the form

$$\delta = \tfrac{1}{2} c_{ij}^k a^i a^j a_k,$$

for coefficients $c_{ij}^k$ such that $c_{ij}^k + c_{ji}^k = 0$. The proof of the following lemma is dual to the proof of Lemma 3.2.

**Lemma 3.3.** *The formula $\delta^2 = 0$ holds if and only if $c_{ij}^k$ are the structure coefficients of a Lie algebra.*

We say that a Lie algebra with cobracket $x_i \mapsto \tfrac{1}{2} f_k^{ij} x_i \wedge x_j$ is a Lie bialgebra if the following identity between the structure coefficients $c_{ij}^k$ and $f_k^{ij}$ is satisfied:

$$c_{mk}^i f_l^{jm} - c_{ml}^i f_k^{jm} - c_{mk}^j f_l^{im} + c_{ml}^j f_k^{im} = c_{kl}^m f_m^{ij}.$$

This is equivalent to supposing that the data $(\mathbf{k} \oplus \mathbf{k}^*, \mathbf{k}, \mathbf{k}^*)$ form a Manin triple, where the inner product on $\mathbf{k} \oplus \mathbf{k}^*$ is induced by the duality between $\mathbf{k}$ and $\mathbf{k}^*$.

A cocharacter of a Lie coalgebra $\mathbf{k}$ is an element $x = p^i a_i \in \mathbf{k}$ lying in the kernel of the cobracket of $\mathbf{k}$.

**Proposition 3.4.** *There is a bijection between structures of a Lie bialgebra on $\mathbf{k}$ together with a cocharacter $x \in \mathbf{k}$ and differential Batalin-Vilkovisky structures on the algebra $\Lambda \mathbf{k}^*$.*

*Proof.* Write $\Delta = \tfrac{1}{2} f_k^{ij} a^k a_i a_j + p^i a_i$ and $\delta = \tfrac{1}{2} c_{ij}^k a^i a^j a_k$. We have

$$\Delta \delta + \delta \Delta = \big( - f_k^{im} c_{lm}^j + \tfrac{1}{4} f_m^{ij} c_{kl}^m \big) a_i a_j a^k a^l + \tfrac{1}{2} f_k^{il} c_{jl}^k (a^j a_i - a_i a^j) + p^i c_{ij}^k a^j a_k.$$

This is a first-order differential operator if and only if the first terms cancel, or equivalently if $\mathbf{k}$ is a Lie bialgebra. $\square$

We will denote the operator $\delta\Delta + \delta\Delta$ by $L$. Note that if $\mathbf{k}$ is a Lie bialgebra,

$$L = D + \mathrm{ad}(x) + \tfrac{1}{2}\mathrm{Tr}(D) = -\mathrm{ad}(\rho_+ - x) - \mathrm{ad}^*(\rho_-) + \tfrac{1}{2}\mathrm{Tr}(D),$$

where $D$ is the derivation $D_i^j a^i a_j$ of $\mathbf{k}$ introduced in the last section and $x = p^i x_i$.

The following dictionary allows us to interpret the structure of a differential Batalin-Vilkovisky algebra on $\Lambda\mathbf{k}^*$ and the $N=2$ superconformal algebra associated to a Manin triple:

| | |
|---|---|
| T | $L$ |
| $\mathsf{G}^+$ | $\delta$ |
| $\mathsf{G}^-$ | $\Delta$ |
| J | degree |

Let $(A, \Delta, \delta)$ be a differential Batalin-Vilkovisky algebra such that the operator $L = \delta\Delta + \Delta\delta$ is semi-simple: the vector space $A$ decomposes into a sum of eigenspaces $A_\lambda$ on which $L$ equals $\lambda$. The basic complex of $A$ is the subspace $A_{\mathrm{basic}}$ on which

$$\Delta a = \Delta\delta a = 0.$$

This is a subcomplex for the differential $\delta$, whose cohomology is called the basic cohomology of $A$. As is explained in the first section of [9], the equivariant cohomology $H_{S^1}^\bullet(A)$ is the derived functor of the basic cohomology. It is defined by replacing $A$ by the differential Batalin-Vilkovisky algebra $A[\Omega, \omega]$, of graded polynomials in a variable $\Omega$ of degree 2 and a variable $\omega$ of degree 1, with operators $\Delta_{\mathrm{tot}}$ and $\delta_{\mathrm{tot}}$

$$\Delta_{\mathrm{tot}} = \frac{\partial}{\partial\omega} + \Delta, \qquad \delta_{\mathrm{tot}} = \Omega\frac{\partial}{\partial\omega} + \delta.$$

Equivalently, the equivariant cohomology is the cohomology of the space of polynomials $A_0[\Omega]$ with respect to the differential $\delta - \Omega\Delta$, where $A_0$ is the kernel of $L$. This construction is the analogue in finite dimensions of the coupling of an $N = 2$ superconformal model with topological gravity.

If $(\mathbf{k}\oplus\mathbf{k}^*, \mathbf{k}, \mathbf{k}^*)$ is a Manin pair, so that $\mathbf{k}^*$ is no longer assumed to be a Lie subalgebra, we say that $\mathbf{k}$ is a quasi-Lie bialgebra: this notion is due to Drinfeld [5]. The cobracket $x_i \mapsto \tfrac{1}{2}f_k^{ij}x_i \wedge x_j$ no longer satisfies the Jacobi relation, and hence the operator $\Delta^2$ does not vanish: rather, we have the formula

$$\Delta^2 + \delta\phi + \phi\delta = 0,$$

where $\phi = \tfrac{1}{6}\phi^{ijk}a_i a_j a_k$. We also have the formulas $\Delta\phi + \phi\Delta = \phi^2 = 0$; we call this structure a quasi-Batalin-Vilkovisky algebra. Note that on the cohomology of $A$ with respect to the differential $\delta$, the operator $\Delta$ induces a Batalin-Vilkovisky structure, since $\Delta^2$ is chain equivalent to 0.

It makes no sense to define the basic cohomology of a quasi-Batalin-Vilkovisky algebra. However, if the action of $L$ is semisimple, we may still define the equivariant cohomology, as the basic cohomology of $A[\omega, \Omega]$ with operators

$$\Delta_{\text{tot}} = \frac{\partial}{\partial \omega} + \Delta - \omega \Delta^2 - \Omega \phi, \qquad \delta_{\text{tot}} = \Omega \frac{\partial}{\partial \omega} + \delta.$$

It may be shown that the equivariant cohomology is the cohomology of the space of polynomials $A_0[\Omega]$ with respect to the differential $\delta - \Omega \Delta + \Omega^2 \phi$, where $A_0$ is the kernel of $L$. We will see in Section 5 that this is a model for the coupling of an $N = 1\frac{1}{2}$ superconformal field theory to topological gravity, and that the coupled theory actually has full $N = 2$ superconformal symmetry.

## 4. A family of $N = 2$ superconformal field theories

Let $(\mathfrak{g}, \mathfrak{g}_+, \mathfrak{g}_-)$ be a Manin pair, with dual bases $x_i$ of $\mathfrak{g}_+$ and $x^i$ of $\mathfrak{g}_-$. We may rewrite the $N = 1$ superconformal model associated to $\mathfrak{g}$ in this basis. The Kac-Moody currents $\{J_i(z), J^i(z)\}$ have operator products

$$J_i(z) \cdot J_j(w) = -\frac{\frac{1}{2}\langle x_i, x_j \rangle}{(z-w)^2} + \frac{c_{ij}^k J_k(w)}{z-w},$$

$$J_i(z) \cdot J^j(w) = \frac{\delta_i^j - \frac{1}{2}\langle x_i, x^j \rangle}{(z-w)^2} + \frac{f_i^{jk} J_k(w) + c_{ki}^j J^k(w)}{z-w},$$

$$J^i(z) \cdot J^j(w) = -\frac{\frac{1}{2}\langle x^i, x^j \rangle}{(z-w)^2} + \frac{f_k^{ij} J^k(w) - \phi^{ijk} J_k(w)}{z-w},$$

while the free fermionic fields $a_i(z)$ and $a^i(z)$ have operator products

$$a_i(z) \cdot a^j(w) \sim \frac{\delta_i^j}{z-w},$$
$$a_i(z) \cdot a_j(w) \sim a^i(z) \cdot a^j(w) \sim 0.$$

The composite currents $I_i(z)$ and $I^i(z)$ are given by the formulas

$$I_i = J_i - c_{ij}^k a^j a_k - \tfrac{1}{2} f_i^{jk} a_j a_k,$$
$$I^i = J^i - f_k^{ij} a_j a^k - \tfrac{1}{2} c_{jk}^i a^j a^k + \tfrac{1}{2} \phi^{ijk} a_j a_k.$$

If $\alpha = p^i x_i + q_i x^i \in \mathfrak{g}_0$, the fields $(\alpha_\pm, I)$ are given by the formulas

$$(\alpha_+, I) = p^i(J_i - c_{ij}^k a^j a_k), \quad (\alpha_-, I) = q_i(J^i - f_k^{ij} a_j a^k).$$

For example, these formulas apply if $\alpha = \rho$.

The field $\mathsf{G}$ of the Kac-Todorov model may decomposed into a sum of three fermionic fields $\mathsf{G}^+ + \mathsf{G}^- + \Phi$, with ghost number repsectively $+1$, $-1$ and $-3$, and given by the formulas

$$\mathsf{G}^+ = J_i a^i - \tfrac{1}{2} c_{ij}^k a^i a^j a_k,$$
$$\mathsf{G}^- = J^i a_i - \tfrac{1}{2} f_k^{ij} a_i a_j a^k,$$
$$\Phi = \tfrac{1}{6} \phi^{ijk} a_i a_j a_k.$$

**Lemma 4.1.**

$$\mathsf{G}^+(z)\cdot\mathsf{G}^-(w) \sim \frac{\tfrac{1}{2}\dim\mathfrak{g} - (\rho,\rho)}{(z-w)^3} + \frac{a^i(w)a_i(w) + (\rho, I(w))}{(z-w)^2}$$
$$+ \frac{\tfrac{1}{2}(J(w), J(w)) + \partial a^i(w)a_i(w) + \tfrac{1}{2}(\rho, \partial I(w))}{z-w},$$

$$\mathsf{G}^+(z)\cdot\mathsf{G}^+(w) \sim 0,$$

$$\mathsf{G}^-(z)\cdot\mathsf{G}^-(w) \sim -\frac{\phi^{ijk}(J_i(w) - c_{il}^m a^l(w)a_m(w))a_j(w)a_k(w)}{z-w} - \frac{c_{kl}^i \phi^{jkl}\partial a_i(w)a_j(w)}{z-w}$$

*Proof.* (1) To calculate $\mathsf{G}^+(z)\cdot\mathsf{G}^-(w)$, we first observe that

$$J_i(z)a^i(z)\cdot J^j(w)a_j(w) \sim \frac{A_3}{(z-w)^3} + \frac{A_2}{(z-w)^2} + \frac{A_1 + B_1}{(z-w)^2},$$

where

$$A_3 = \tfrac{1}{2}(\dim\mathfrak{g} - \langle x_i, x^i\rangle) = \tfrac{1}{2}\dim\mathfrak{g} + \tfrac{3}{2}\operatorname{Tr}(D),$$
$$A_2 = (\delta_i^j - \tfrac{1}{2}\langle x_i, x^j\rangle)a^i(w)a_j(w) + (\rho, J(w)),$$
$$= a^i(w)a_i(w) + (-\tfrac{1}{2}D_i^j + A_i^j)a^i(w)a_j(w) + (\rho, I(w)),$$
$$A_1 = (\delta_i^j - \tfrac{1}{2}\langle x_i, x^j\rangle)\partial a^i(w)a_j(w) + J_i(w)J^i(w),$$
$$= \partial a^i(w)a_i(w) - \tfrac{1}{2}D_i^j a^i(w)\partial a_j(w) + A_i^j \partial a^i(w)a_j(w)$$
$$+ \tfrac{1}{2}\Big(J_i(w)J^i(w) + J^i(w)J_i(w) + (\rho, \partial I(w))\Big),$$
$$B_1 = f_i^{jk} J_k(w)a^i(w)a_j(w) + c_{ki}^j J^k(w)a^i(w)a_j(w).$$

Here, we have made use of the formulas

$$J_i(z)J^i(z) = \tfrac{1}{2}\Big((J(z), J(z)) + (\rho, \partial J(z))\Big)$$

and

$$(\rho, \partial I(z)) = (\rho, \partial J(z)) + D_i^j a^i(z)a_j(z).$$

The term $B_1$ is cancelled by the operator products

$$J_i(z)a^i(z)\cdot(-\tfrac{1}{2}f_l^{jk}a_j(w)a_k(w)a^l(w)) \sim -\frac{f_i^{jk}J_k(w)a^i(w)a_j(w)}{z-w},$$
$$(-\tfrac{1}{2}c_{jk}^l a^j(z)a^k(z)a_l(z))\cdot J^i(w)a_i(w) \sim -\frac{c_{kj}^i J^k(w)a^i(w)a_j(w)}{z-w}.$$

The calculation of the operator product $\mathsf{G}^+(z)\cdot\mathsf{G}^-(w)$ is completed by the formula

$$(-\tfrac{1}{2}c_{ij}^k a^i(z)a^j(z)a_k(z))\cdot(-\tfrac{1}{2}f_r^{pq}a_p(w)a_q(w)a^r(w)) \sim$$
$$-\frac{\tfrac{1}{2}\operatorname{Tr}(D)}{(z-w)^3} + \frac{(\tfrac{1}{2}D_i^j - A_i^j)a^i(w)a_j(w)}{(z-w)^2} + \frac{\tfrac{1}{2}D_i^j a^i(w)\partial a_j(w)}{z-w} - \frac{A_i^j \partial a^i(w)a_j(w)}{z-w}.$$

(2) To calculate $\mathsf{G}^+(z) \cdot \mathsf{G}^+(w)$, we first observe that

$$J_i(z)a^i(z) \cdot J_j(w)a^j(w) \sim \frac{A}{z-w} + \frac{B}{z-w},$$

where $A = c_{ij}^k J_k(w)a^i(w)a^j(w)$ and $B = -\frac{1}{2}\langle x_i, x_j\rangle \partial a^i(w)a^j(w)$. The term $A$ is cancelled by the operator products

$$J_i(z)a^i(z) \cdot (-\tfrac{1}{2}c_{jk}^l a^j(w)a^k(w)a_l(w)) \sim -\tfrac{1}{2}\frac{c_{jk}^i J_i(w)a^j(w)a^k(w)}{z-w},$$

$$(-\tfrac{1}{2}c_{jk}^l a^j(z)a^k(z)a_l(z)) \cdot J_i(w)a^i(w) \sim -\tfrac{1}{2}\frac{c_{jk}^i J_i(w)a^j(w)a^k(w)}{z-w}.$$

The term $B$ is cancelled by the operator product

$$(-\tfrac{1}{2}c_{jk}^l a^j(z)a^k(z)a_l(z)) \cdot (-\tfrac{1}{2}c_{jk}^l a^j(w)a^k(w)a_l(w)) \sim \tfrac{1}{2}\frac{\langle x_i, x_j\rangle \partial a^i(w)a^j(w)}{z-w}.$$

(3) To calculate $\mathsf{G}^-(z) \cdot \mathsf{G}^-(w)$, we first observe that

$$J^i(z)a_i(z) \cdot J^j(w)a_j(w) \sim \frac{A}{z-w} + \frac{B}{z-w} + \frac{C}{z-w},$$

where $A = -\phi^{ijk}J_i(w)a_j(w)a_k(w)$, $B = f_k^{ij}J^k(w)a_i(w)a_j(w)$, and $C = -\frac{1}{2}\langle x^i, x^j\rangle \partial a_i(w)a_j(w)$. The term $B$ is cancelled by the operator products

$$J^l(z)a_l(z) \cdot (-\tfrac{1}{2}f_k^{ij}a_i(w)a_j(w)a^k(w)) \sim -\tfrac{1}{2}\frac{f_k^{ij}J^k(w)a_i(w)a_j(w)}{z-w},$$

$$(-\tfrac{1}{2}f_k^{ij}a_i(z)a_j(z)a^k(z)) \cdot J^l(w)a_l(w) \sim -\tfrac{1}{2}\frac{f_k^{ij}J^k(w)a_i(w)a_j(w)}{z-w}.$$

Next observe that

$$(-\tfrac{1}{2}f_k^{ij}a_i(z)a_j(z)a^k(z)) \cdot (-\tfrac{1}{2}f_n^{lm}a_l(w)a_m(w)a^n(w)) \sim \frac{D}{z-w} + \frac{E}{z-w},$$

where $D = f_m^{ij}f_l^{mk}a_i(w)a_j(w)a_k(w)a^l(w)$ and $E = f_l^{ik}f_k^{jl}\partial a_i(w)a_j(w)$. The proof is completed using the formulas

$$f_m^{ij}f_l^{mk} + f_m^{jk}f_l^{mi} + f_m^{ki}f_l^{mj} = c_{lm}^i\phi^{jkm} + c_{lm}^j\phi^{kim} + c_{lm}^k\phi^{ijm},$$

$$\langle x^i, x^j\rangle = 2f_l^{ik}f_k^{jl} + c_{kl}^i\phi^{jkl} + c_{kl}^j\phi^{ikl}. \quad \square$$

Motivated by this lemma, we introduce fields

$$\mathsf{J} = a^i a_i + (\rho, I),$$
$$\mathsf{T} = \tfrac{1}{2}\Big(J_i J^i + J^i J_i + \partial a^i a_i - a^i \partial a_i\Big),$$
$$\mathsf{F} = \tfrac{1}{2}\phi^{ijk}(J_i - c_{il}^m a^l a_m)a_j a_k + \tfrac{1}{2}c_{kl}^i \phi^{jkl}\partial a_i a_j.$$

Then the above lemma shows that

$$\mathsf{G}^+(z) \cdot \mathsf{G}^-(w) \sim \frac{\frac{1}{2}\dim \mathfrak{g} - (\rho,\rho)}{(z-w)^3} + \frac{\mathsf{J}(w)}{(z-w)^2} + \frac{\mathsf{T}(w) + \frac{1}{2}\mathsf{J}(w)}{z-w},$$

$$\mathsf{G}^-(z) \cdot \mathsf{G}^-(w) \sim -\frac{2\mathsf{F}(w)}{z-w}.$$

It is easy to see that

$$\mathsf{G}^+(z) \cdot \Phi(w) \sim \frac{\mathsf{F}(w)}{z-w}.$$

To complete the verification of the relations of the $N = 1\frac{1}{2}$ superconformal algebra, we must show that the fields $\mathsf{G}^\pm$ and $\Phi$ have charges $\pm 1$ and $-3$. The proof uses the following lemma.

**Lemma 4.2.** *If $\alpha \in \mathfrak{g}_0$, the superfields $\Omega^\pm(Z) = (\alpha_\mp, \mathcal{J}(Z))$ satisfy the hypotheses of Proposition 1.4.*

*Proof.* Let us start with the case of $\Omega^+(Z) = A^+(z) + \theta B^+(z)$, where

$$A^+ = q_i a^i, \quad B^+ = q_i(J^i - f^{ij}_k a_j a^k).$$

It may be shown by the same methods as in the proof of Proposition 1.2 that it suffices to prove the following operator products:

$$\mathsf{G}^-(z) \cdot A^+(w) \sim \frac{B^+(w)}{z-w},$$

$$\mathsf{G}^+(z) \cdot B^+(w) \sim \frac{A^+(w)}{(z-w)^2} + \frac{\partial A^+(w)}{z-w},$$

$$\mathsf{G}^+(z) \cdot A^+(w) \sim \Phi(z) \cdot A^+(w) \sim 0.$$

The vanishing of $q_i c^i_{jk}$ shows that $\mathsf{G}^+(z) \cdot A^+(w) \sim 0$, while the vanishing of $q_i \phi^{ijk}$ shows that $\Phi(z) \cdot A^+(w) \sim 0$. The formula

$$\mathsf{G}^-(z) \cdot a^i(w) \sim \frac{J^i(w) - f^{ij}_k a_j(w) a^k(w)}{z-w}$$

shows that $\mathsf{G}^-(z) \cdot A^+(w) \sim B^+(w)/(z-w)$.

To calculate $B^+(z) \cdot \mathsf{G}^+(w)$, observe that

$$q_i J^i(z) \cdot J_j(w) a^j(w) \sim \frac{A_2}{(z-w)^2} + \frac{A_1 + B_1}{z-w},$$

where $A_2 = (\delta^i_j - \frac{1}{2}\langle x^i, x_j\rangle) q_i a^j(w)$, $D = c^i_{jk} q_i a^j(w) J^k(w)$ and $E = -f^{ik}_j q_i a^j(w) J_k(w)$. The term $B_1$ is cancelled by the operator product

$$q_i(-f^{ik}_j a_k(z) a^j(z)) \cdot J_l(w) a^l(w) \sim \frac{f^{ik}_j q_i a^j(w) J_k(w)}{z-w}.$$

The calculation of $\mathsf{G}^+(z) \cdot B^+(w)$ is completed by the operator product

$$q_i(-f_k^{ij}a_j(z)a^k(z)) \cdot (-\tfrac{1}{2}c_{lm}^n a^l(w)a^m(w)a_n(w)) \sim \frac{\tfrac{1}{2}\langle x_j, x^i\rangle q_i a^j(w)}{(z-w)^2}.$$

The case of $\Omega^-(Z) = A^-(z) + \theta B^-(z)$, where

$$A^- = p^i a_i, \quad B^- = p^i(J_i - c_{ij}^k a^j a_k),$$

is proved in a similar way, by proving the formulas

$$\mathsf{G}^+(z) \cdot A^-(w) \sim \frac{B^-(w)}{z-w},$$

$$\mathsf{G}^-(z) \cdot B^-(w) \sim \frac{A^-(w)}{(z-w)^2} + \frac{\partial A^-(w)}{z-w},$$

$$\mathsf{G}^-(z) \cdot A^-(w) \sim \Phi(z) \cdot A^-(w) \sim 0. \quad \square$$

Using this lemma, we see that

$$\mathsf{G}^+(z) \cdot (\rho, I(w)) \sim \mathsf{G}^+(z) \cdot (\rho_+, I(w)) + \mathsf{G}^+(z) \cdot (\rho_-, I(w))$$
$$\sim \frac{(\rho_+, a(w))}{(z-w)^2} + \frac{(\rho_+, \partial a(w))}{z-w},$$

$$\mathsf{G}^-(z) \cdot (\rho, I(w)) \sim \mathsf{G}^-(z) \cdot (\rho_+, I(w)) + \mathsf{G}^-(z) \cdot (\rho_-, I(w))$$
$$\sim \frac{(\rho_-, a(w))}{(z-w)^2} + \frac{(\rho_-, \partial a(w))}{z-w}.$$

Furthermore,

$$a^i(z)a_i(z) \cdot \mathsf{G}^+(w) \sim \frac{c_{ij}^j a^i(w)}{(z-w)^2} + \frac{c_{ij}^j \partial a^i(w)}{z-w} \sim \frac{(\rho_+, a(w))}{(z-w)^2} + \frac{(\rho_+, \partial a(w))}{z-w},$$

$$a^i(z)a_i(z) \cdot \mathsf{G}^-(w) \sim \frac{f_j^{ij} a_i(w)}{(z-w)^2} + \frac{f_j^{ij} \partial a_i(w)}{z-w} \sim \frac{(\rho_-, a(w))}{(z-w)^2} + \frac{(\rho_-, \partial a(w))}{z-w}.$$

Combining these formulas, we see that $\mathsf{J}(z) \cdot \mathsf{G}^\pm(w) \sim \pm\mathsf{G}^\pm(w)/(z-w)$.

To complete the verification of the $N = 1\tfrac{1}{2}$ superconformal algebra, we must check that $\mathsf{G}^-(z) \cdot \Phi(w) \sim 0$, which is a consequence of the formula

$$f_m^{il}\phi^{jkm} + f_m^{jl}\phi^{klm} + f_m^{kl}\phi^{ijm} = 0,$$

and that

$$\mathsf{J}(z) \cdot \Phi(w) \sim -\frac{3\Phi(w)}{z-w},$$

which follows from the fact that $(\rho, I(z)) \cdot \Phi(w) \sim 0$.

Having constructed an action of the $N = 1\tfrac{1}{2}$ superconformal algebra, it is simple to construct others: we simply apply Proposition 1.4 with

$$\Omega^\pm(z) = (\alpha_\mp, \mathcal{I}(z)),$$

where $\alpha = \alpha_+ + \alpha_-$ is an element of $\mathfrak{g}_0$. This has the effect of modifying the fields to

$$\begin{aligned}
\mathsf{T} &= \tfrac{1}{2}\bigl(J_i J^i + J^i J_i + \partial a^i a_i - a^i \partial a_i + (\alpha, \partial I)\bigr), \\
\mathsf{G}^+ &= J_i a^i - \tfrac{1}{2} c^k_{ij} a^i a^j a_k + q_i \partial a^i, \\
\mathsf{G}^- &= J^i a_i - \tfrac{1}{2} f^{ij}_k a_i a_j a^k + p^i \partial a_i, \\
\mathsf{J} &= a^i a_i + (\rho + \bar\alpha, I),
\end{aligned}$$

while $\Phi$ and $\mathsf{F}$ are unchanged. The central charge becomes $d = \tfrac{1}{2} \dim \mathfrak{g} - (\rho, \rho) - (\alpha, \alpha)$.

There is an especially natural choice for $\alpha$, namely $\alpha = -\bar\rho$. With this value of $\alpha$, the field $\mathsf{J}$ takes the especially simple form $a^i a_i$ of the ghost number, while the central charge becomes simply $d = \tfrac{1}{2} \dim \mathfrak{g}$. It is natural to restrict $\alpha$ to lie in the lattice for which the zero-mode of the field $\mathsf{J}$ has integral eigenvalues: this is equivalent to the zero-mode of the field $(\rho + \bar\alpha, I(z))$ having integral eigenvalues.

## 5. Coupling to topological gravity

There is a modification of the $N = 1\tfrac{1}{2}$ superconformal algebra with twisted boundary conditions, parametrized by an angle $\theta$, in which the fields $\mathsf{G}^\pm(z)$ and $\Phi(z)$ satisfy the boundary conditions

$$\begin{aligned}
\mathsf{G}^\pm(e^{2\pi i} z) &= e^{\pm i\theta} \mathsf{G}^\pm(z), \\
\Phi(e^{2\pi i} z) &= e^{-3i\theta} \Phi(z).
\end{aligned}$$

We may realize the $N = 1\tfrac{1}{2}$ superconformal algebra with these boundary conditions by using the construction of the last section but with fermions having boundary conditions

$$a^k(e^{2\pi i} z) = e^{i\theta} a^k(z) \quad, \quad a_k(e^{2\pi i} z) = e^{-i\theta} a_k(z).$$

Of these boundary conditions, only the cases where $e^{i\theta} = \pm 1$ have an underlying $N = 1$ superconformal symmetry generated by the field $\mathsf{G}(z) = \mathsf{G}^+(z) + \mathsf{G}^-(z) + \Phi(z)$: the choice of periodic boundary conditions $e^{i\theta} = 1$ goes by the name of the Ramond sector, while the choice of anti-periodic boundary conditions $e^{i\theta} = -1$ is called the Neveu-Schwarz sector.

The twisting construction of Witten and Eguchi-Yang is a way of constructing a topological field theory in its Ramond (Neveu-Schwarz) sector from an $N = 1\tfrac{1}{2}$ theory in its Neveu-Schwarz (Ramond) sector. The Hilbert spaces of the two theories, are identical, and all that is changed is the stress-energy tensor $\mathsf{T}(z)$, which is replaced by its twist $\mathsf{T}_{\mathrm{top}}(z) = \mathsf{T}(z) + \tfrac{1}{2} \partial \mathsf{J}(z)$. This has the effect of modifying the conformal dimensions of the fields in the theory, subtracting from them half of their U(1) charge. In particular, the field $\mathsf{G}^+(z)$ now has conformal dimension 1. In the Ramond sector of the topological theory, this allows us to think of the zero mode $\mathsf{Q}$ of the field $\mathsf{G}^+$ as a differential on the space of states of the theory, and the zero mode of the field $\mathsf{J}$ as a degree operator. (The operator $\mathsf{Q}$ may be identified with the mode $\mathsf{G}^+_{1/2}$ of the untwisted $N = 2$ theory.)

The operator $\mathsf{Q}$ raises degree by 1, and its cohomology is by definition the space of physical states of the topological field theory. By the formula

$$\mathsf{T}_{\mathrm{top}}(z) = [\mathsf{Q}, \mathsf{G}^-(z)],$$

we see that the Virasoro algebra acts trivially on this cohomology space, explaining why this model is called a topological field theory. In the model of the last section, this differential

$$\mathsf{Q} = \mathrm{Res}\left(J_i a^i - \tfrac{1}{2} c^k_{ij} a^i a^j a_k\right)$$

is the BRS operator which calculates the semi-infinite cohomology for the projective action of the infinite dimensional Lie algebra $\widehat{\mathfrak{g}_+}$ acting through the currents $J_i(z)$. Note that this differential is independent of $\alpha \in \mathfrak{g}_0$, and hence so is the physical Hilbert space. By contrast, the zero mode of $\mathsf{J}$, which defines the grading, and the zero-mode of the spin 2 field $\mathsf{G}^-$ do depend on $\alpha$.

In [9], we have described the coupling of a topological field theory associated to an $N = 2$ superconformal field theory to topological gravity. This may be generalized to the case of an $N = 1\tfrac{1}{2}$ superconformal action. The semi-infinite Weil complex $W$ is the Fock space associated to a pair $\{\gamma(z), \beta(z)\}$ of free bosonic fields and a pair $\{c(z), b(z)\}$ of free fermionic fields, with operator products

$$b(w) \cdot c(z) \sim \beta(z) \cdot \gamma(w) \sim \frac{1}{z-w},$$

while $c(z) \cdot c(w) \sim b(z) \cdot b(w) \sim \gamma(z) \cdot \gamma(w) \sim \beta(z) \cdot \beta(w) \sim 0$. The vacuum $|-1\rangle$ of $W$ is characterized by its being annihilated by the modes

$$\{b_n, \beta_n \mid n \geq -1\} \cup \{c_n, \gamma_n \mid n > 1\};$$

this Fock space is called the $(-1)$-picture.

The couple theory is the tensor product of the highest weight module $\mathcal{E}$ of the $N = 1\tfrac{1}{2}$ superconformal algebra with the semi-infinite Weil complex $W$, and has an action of the $N = 2$ superconformal algebra, with central charge $d - 3$, given by the formulas

$$\begin{aligned}
\mathbb{T}_{\mathrm{top}} &= 2\partial cb + c\partial b + 2\partial \gamma \beta + \gamma \partial \beta + \mathsf{T}_{\mathrm{top}}, \\
\mathbb{G}^+ &= \gamma b + \mathsf{G}^+, \\
\mathbb{G}^- &= b + 2\partial c\beta + c\partial \beta + \mathsf{G}^- + c\mathsf{F} - \gamma\Phi, \\
\mathbb{J} &= cb + 2\gamma\beta + \mathsf{J}.
\end{aligned}$$

It is interesting that on coupling to topological gravity, we recover a full $N = 2$ superconformal symmetry, despite only having $N = 1\tfrac{1}{2}$ superconformal symmetry in the matter sector.

Suppose that $\mathcal{E}$ is the vector space of states of an $N = 1\tfrac{1}{2}$ superconformal field theory: thus, it has two commuting actions of the $N = 1\tfrac{1}{2}$ superconformal algebra, corresponding to chiral and anti-chiral sectors of the conformal field theory. The cohomology $H^\bullet(\mathcal{E})$ of the differential $Q + \bar{Q}$ is a commutative algebra, with respect to the fusion product, while the operator $G_0^- - \bar{G}_0^-$ induces an operator $\Delta$ on $H^\bullet(\mathcal{E})$ giving it the structure of a Batalin-Vilkovisky algebra.

Define the equivariant cohomology $H^\bullet_{S^1}(\mathcal{E})$ of $\mathcal{E}$ to be the semi-relative cohomology of the coupled theory $W \otimes \bar{W} \otimes \mathcal{E}$, that is, the cohomology of the BRS operator $\mathbb{Q} + \bar{\mathbb{Q}}$ on the kernel of the difference of zero-modes $\mathbb{G}_0^- - \bar{\mathbb{G}}_0^-$. This is the space of physical states of the topological field theory underlying $\mathcal{E}$ to topological gravity. The equivariant cohomology $H^\bullet_{S^1}(\mathcal{E})$ is a gravity algebra, just as in the case of an $N = 2$ superconformal field theory.

The equivariant cohomology of the model introduced in the last section is independent of $\alpha_- \in \mathfrak{g}_0$, and we conjecture that it is independent of $\alpha_+$ as well, in the sense that there is a flat (or, possibly, projectively flat) connection on the bundle over $\mathfrak{g}_+ \cap \mathfrak{g}_0$ whose fibre at $\alpha_+$ is the equivariant cohomology of the corresponding model.

Define the field
$$\mathbb{K} = c(\partial c \beta + \mathsf{G}^- - \gamma \Phi),$$
of conformal dimension 1 and charge 0; thus, the fields $\mathbb{T}_{\text{top}}$ and $\mathbb{J}$ are invariant under the operation $e^{\text{ad}(\mathbb{K})}$ of conjugation by the exponential of the zero mode of $\mathbb{K}$.

**Proposition 5.1.**
$$\begin{aligned} e^{\text{ad}(\mathbb{K})}\mathbb{G}^+ &= c\partial cb + c(2\partial\gamma\beta + \gamma\partial\beta + \mathsf{T}_{\text{top}}) \\ &\quad + \mathsf{G}^+ + \gamma(b - \mathsf{G}^-) + \gamma^2\Phi - \partial(c\mathbb{J}) + \tfrac{d-3}{2}\partial^2 c, \\ e^{\text{ad}(\mathbb{K})}\mathbb{G}^- &= b. \end{aligned}$$

*Proof.* The formula for $e^{\text{ad}(\mathbb{K})}\mathbb{G}^+$ follows from the formulas
$$\begin{aligned} \text{ad}(\mathbb{K})\mathbb{G}^+ &= c\partial cb + c(2\partial\gamma\beta + \gamma\partial\beta + \mathsf{T}_{\text{top}}) \\ &\quad + \mathsf{G}^+ - \gamma\mathsf{G}^- + \gamma^2\Phi - \partial(c\mathbb{J}) + \tfrac{d-3}{2}\partial^2 c - c\gamma\mathsf{F}, \\ \tfrac{1}{2}\text{ad}(\mathbb{K})^2\mathbb{G}^+ &= c\gamma\mathsf{F} - c\partial c\gamma\Phi, \\ \tfrac{1}{6}\text{ad}(\mathbb{K})^3\mathbb{G}^+ &= c\partial c\gamma\Phi, \quad \text{ad}(\mathbb{K})^4\mathbb{G}^+ = 0. \end{aligned}$$

The formula for $e^{\text{ad}(\mathbb{K})}\mathbb{G}^-$ follows from the formulas
$$\begin{aligned} \text{ad}(\mathbb{K})\mathbb{G}^+ &= -2\partial c\beta - c\partial\beta - \mathsf{G}^- - 2c\mathsf{F} + \gamma\Phi + c\partial c\Phi, \\ \tfrac{1}{2}\text{ad}(\mathbb{K})^2\mathbb{G}^+ &= c\mathsf{F} - 2c\partial c\Phi, \\ \tfrac{1}{6}\text{ad}(\mathbb{K})^3\mathbb{G}^+ &= c\partial c\Phi, \quad \text{ad}(\mathbb{K})^4\mathbb{G}^+ = 0. \quad \square \end{aligned}$$

We may now prove the following extension of Theorem 2.2 of [9].

**Theorem 5.2.** *The equivariant cohomology $H^\bullet_{S^1}(\mathcal{E})$ is naturally isomorphic to the cohomology of the complex $\mathcal{E}_0[\Omega]$ of polynomials in a variable $\Omega$ of ghost number 2 with* *coefficients in the kernel $\mathcal{E}_0$ of the rotation operator $(\mathsf{T}_{\mathrm{top}})_0 - (\bar{\mathsf{T}}_{\mathrm{top}})_0$, with respect to the differential*

$$(\mathsf{Q} + \bar{\mathsf{Q}}) - \Omega(\mathsf{G}_0^- - \bar{\mathsf{G}}_0^-) + \Omega^2(\Phi_0 + \bar{\Phi}_0).$$

*Proof.* The operator

$$kc_{-k}\beta_k + \mathsf{G}_0^-$$

has degree $-1$, and commutes with the operators $(\mathbb{T}_{\mathrm{top}})_0$ and $\mathbb{G}_0^-$; in proving this, we use the formula $[\mathsf{G}_n^-, \mathsf{F}_m] = (2n - m)\Phi_{n+m}$. Thus, it restricts to a map of the basic complex $(W \otimes \bar{W} \otimes \mathcal{E})_{\mathrm{basic}}$ to itself. Thus, we may calculate the equivariant cohomology by restricting attention to the kernel of the commutator

$$[\mathbb{G}_0^+ + \bar{\mathbb{G}}_0^+, kc_{-k}\beta_k + \mathsf{G}_0^-] = k(c_{-k}b_k + \gamma_{-k}\beta_k) + (\mathsf{T}_{\mathrm{top}})_0,$$

acting on the basic subcomplex. Taking $k = 0$, we see that we may suppose that $(\mathsf{T}_{\mathrm{top}})_0 = 0$. For $k < -1$, the operators $c_{-k}b_k$ and $\gamma_{-k}\beta_k$ are negative, and we may suppose they each vanish. For $k > 0$ or $k = -1$, the operators $c_{-k}b_k$ and $\gamma_{-k}\beta_k$ are positive, and we may again suppose they vanish. Similar arguments apply in the anti-chiral sector, so we may calculate the equivariant cohomology on the subcomplex spanned by the vectors

$$\left\{c_0^m \bar{c}_0^{\bar{m}} \gamma_0^n \bar{\gamma}_0^{\bar{n}} |-1\rangle \mid m, \bar{m} \in \{0, 1\}, n, \bar{n} \geq 0\right\} \otimes \{v \in \mathcal{E} \mid (\mathsf{T}_{\mathrm{top}})_0 v = (\bar{\mathsf{T}}_{\mathrm{top}})_0 v = (\mathsf{G}_0^- - \bar{\mathsf{G}}_0^-)v = 0\}.$$

The rest of the proof is as in [9]. □

## 6. Twisting

Given an element $R \in \Lambda^2 \mathfrak{g}_+$, we have shown in Section 2 how to twist a Manin pair $(\mathfrak{g}, \mathfrak{g}_+, \mathfrak{g}_-)$, to obtain a new Manin pair in which the space $\mathfrak{g}_-$ is replaced by $R \cdot \mathfrak{g}_-$, spanned by the vectors $x^i - R^{ij}x_j$. In this section, we describe how the $N = 1\frac{1}{2}$ superconformal symmetry is modified by twisting.

If $R^{ij}$ is the antisymmetric matrix defining a twist, introduce the operator $\mathsf{R} = \frac{1}{2}R^{ij}a_i a_j$. The following formulas are easily proved:

$$\mathsf{G}^+(z) \cdot \mathsf{R}(w) \sim \frac{\frac{1}{2}c_{ij}^k R^{ij} a_k(w) - q_i R^{ij} a_k(w)}{(z-w)^2} - \frac{R^{ij} a_i(w)(J_j(w) - c_{jk}^l a^k(w)a_l(w))}{z-w}$$
$$+ \frac{\frac{1}{2}c_{ij}^k R^{ij} \partial a_k(w)}{(z-w)^2},$$

$$\mathsf{G}^-(z) \cdot \mathsf{R}(w) \sim -\frac{\frac{1}{2}f_l^{ij} R^{kl} a_i(w) a_j(w) a_k(w)}{z-w}.$$

We modify the fermionic fields in the following way: $\tau_\mathsf{R} \mathsf{G}^+ = \mathsf{G}^+$, while

$$\tau_\mathsf{R} \mathsf{G}^- = \mathsf{G}^- + [\mathsf{G}^+ \mathsf{R}]_1,$$
$$\tau_\mathsf{R} \Phi = \Phi - \tfrac{2}{3}[\mathsf{G}^- \mathsf{R}]_1 + \tfrac{1}{6}[[\mathsf{G}^+ \mathsf{R}]_1 \mathsf{R}]_1.$$

**Proposition 6.1.** *The twisted fields $\tau_R \mathsf{G}^\pm$ and $\tau_R \Phi$ are the fields associated to the twisted Manin pair $(\mathfrak{g}, \mathfrak{g}_+, R \cdot \mathfrak{g}_-)$, with $\alpha_+$ replaced by*

$$\tau_R \alpha_+ = \alpha_+ - \tfrac{1}{2} c_{ij}^k R^{ij} x_k,$$

*while $\alpha_-$ is unchanged. They generate an action of the $N = 1\tfrac{1}{2}$ superconformal algebra if and only if*

$$\operatorname{ad}(\tau_R \alpha_+) R^{lm} + \tfrac{1}{2} f_k^{lm} c_{ij}^k R^{ij} = \operatorname{ad}^*(\alpha_-) R = 0.$$

*The twists of the other fields are given by the formulas*

$$\tau_R \mathsf{J} = \mathsf{J} - [\mathsf{G}^+ [\mathsf{G}^+ R]_2]_1,$$
$$\tau_R \mathsf{T} = \mathsf{T} - \tfrac{1}{2} \partial [\mathsf{G}^+ [\mathsf{G}^+ R]_2]_1,$$
$$\tau_R \mathsf{F} = \mathsf{F} - \tfrac{2}{3} [\mathsf{G}^+ [\mathsf{G}^- R]_1]_1 + \tfrac{1}{6} [\mathsf{G}^+ [[\mathsf{G}^+ R]_1 R]_1]_1.$$

*Proof.* Since twisting sends the coefficients $f_k^{ij}$ to

$$f_k^{ij} - R^{il} c_{kl}^j - R^{lj} c_{kl}^i$$

and the vector $x^i$ to $x^i - R^{ij} x_j$, it twists the field $J^i a_i - \tfrac{1}{2} f_k^{ij} a_i a_j a^k$ to

$$(J^i - R^{ij} J_k) a_i - \tfrac{1}{2} f_k^{ij} a_i a_j a^k + R^{il} c_{kl}^j a_i a_j a_k = J^i a_i - \tfrac{1}{2} f_k^{ij} a_i a_j a^k - R^{ij} a_i (J_j - c_{jk}^l a^k a_l),$$

while leaving the field $q_i \partial a^i$ invariant. This shows that $\mathsf{G}^- + [\mathsf{G}^+ R]_1$ equals the field $\mathsf{G}^-$ associated to the twisted Manin triple $(\mathfrak{g}, \mathfrak{g}_+, R \cdot \mathfrak{g}_-)$, but with $\alpha_+$ replaced by $\tau_R \alpha_+$.

Twisting sends the coefficients $\phi^{ijk}$ to

$$\phi^{ijk} + R^{il} f_l^{jk} - R^{jl} f_l^{ik} - R^{il} R^{jm} c_{lm}^k$$

The formulas

$$(R^{il} f_l^{jk} - R^{jl} f_l^{ik}) a_i a_j a_k = -4[\mathsf{G}^- R]_1,$$
$$R^{il} R^{jm} c_{lm}^k a_i a_j a_k = [[\mathsf{G}^+ R]_1 R]_1$$

show that the field $\tau_R \Phi$ associated to the twisted Manin triple $(\mathfrak{g}, \mathfrak{g}_+, R \cdot \mathfrak{g}_-)$ equals $\Phi - \tfrac{2}{3} [\mathsf{G}^- R]_1 + \tfrac{1}{6} [[\mathsf{G}^+ R]_1 R]_1$.

The twisted field $\tau_R \mathsf{J}$ is given by the formula

$$\tau_R \mathsf{J} = [\mathsf{G}^+ \tau_R \mathsf{G}^-]_2 = \mathsf{J} + [\mathsf{G}^+ [\mathsf{G}^+ R]_1]_2 = \mathsf{J} - [\mathsf{G}^+ [\mathsf{G}^+ R]_2]_1.$$

The twisted stress-energy tensor $\tau_R \mathsf{T}$ is given by the formula

$$\tau_R \mathsf{T} = [\mathsf{G}^+ \tau_R \mathsf{G}^-]_1 - \tfrac{1}{2} \partial \tau_R \mathsf{J} = \mathsf{T} + [\mathsf{G}^+ [\mathsf{G}^+ \mathsf{G}^-]_1]_1 + \tfrac{1}{2} \partial [\mathsf{G}^+ [\mathsf{G}^+ R]_2]_1.$$

However, $[\mathsf{G}^+ [\mathsf{G}^+ \mathsf{G}^-]_1]_1 = 0$, while $\partial [\mathsf{G}^+ [\mathsf{G}^+ R]_2]_1 = [\mathsf{G}^+ \partial [\mathsf{G}^+ R]_2]_1$. The formula for $\tau_R \mathsf{F}$ follows from the fact that $\tau_R \mathsf{F} = [\mathsf{G}^+ \tau_R \Phi]_1$.

The twisted fields $\tau_R \mathsf{G}^\pm$ and $\tau_R \Phi$ generate an action of the $N = 1\tfrac{1}{2}$ superconformal algebra provided the following three conditions hold:

(1) $\tau_R \alpha_+ \perp [R \cdot \mathfrak{g}_-, R \cdot \mathfrak{g}_-]$;

(2) $\alpha_- \perp [\mathfrak{g}_+, \mathfrak{g}_+]$;
(3) $\alpha_- \perp [R \cdot \mathfrak{g}_-, R \cdot \mathfrak{g}_-]$.

The first condition may be rewritten

$$(p^k - \tfrac{1}{2} c^k_{lm} R^{lm})(f^{ij}_k - R^{il} c^j_{kl} + R^{jl} c^i_{kl}) = 0.$$

Since $p^k f^{ij}_k$ vanishes and

$$(p^k - \tfrac{1}{2} c^k_{lm} R^{lm})(R^{il} c^j_{kl} + R^{lj} c^i_{kl}) = \mathrm{ad}(\alpha_+) R,$$

the condition on $\alpha_+$ follows.

The second of these conditions is automatically satisfied, while the third may be rewritten

$$q_k(R^{il} f^{jk}_l - R^{jl} f^{ik}_l - R^{il} R^{jm} c^k_{lm}) = q_k(R^{il} f^{jk}_l - R^{jl} f^{ik}_l) = 0,$$

where we have used the fact that, since $\alpha_- \in \mathfrak{g}_0$, $q_k c^k_{ij} = 0$. But this equation is equivalent to $\mathrm{ad}^*(\alpha_-) R = 0$. $\square$

From this proposition, we see that the twist does not change the BRS operator $\mathsf{Q}$, and only changes the field $\mathsf{J}$ by a BRS commutator $[\mathsf{Q}, [\mathsf{G}^+ \mathsf{R}]_2]$. Thus, the cohomology of the space of physical states, and its grading, is unchanged. Since the field $\mathsf{G}^-$ is changed by a BRS commutator $[\mathsf{Q}, \mathsf{R}]$, the action of the Batalin-Vilkovisky operator $\Delta$ on the cohomology is unchanged.

The equivariant cohomology is also left unchanged by a twist, since the field $\mathbb{G}^+ = \gamma\beta + \mathsf{G}^+$ is unchanged, while $\mathbb{G}^-$ changes by a BRS commutator:

$$\tau_\mathsf{R} \mathbb{G}^- = \mathbb{G}^- + [\mathbb{Q}, \mathsf{R} + \tfrac{2}{3} c[\mathsf{G}^- \mathsf{R}]_1 - \tfrac{1}{6} c[[\mathsf{G}^+ \mathsf{R}]_1 \mathsf{R}]_1].$$

## 7. The $\mathrm{SL}(2)/\mathrm{SL}(2)$ model

In this section, we study two $N = 2$ superconformal actions, associated to the two conjugacy classes of parabolic subalgebras of the Lie algebra $\mathbf{k} = \mathbf{sl}(2)$, $\mathbf{p} = \mathbf{b}$ and $\mathbf{p} = \mathbf{k}$. We will construct an intertwining operator between these two superconformal algebras, or rather between certain $N = 1\tfrac{1}{2}$ twists of them, in the sense of the last section. Aharony et al. [2] have conjectured that similar results hold for any simple Lie algebra $\mathbf{k}$: for example, that for any simple Lie algebra $\mathbf{k}$, there is an explicit intertwining operator between the $N = 1\tfrac{1}{2}$ superconformal action associated to the Manin pair based on $\mathbf{k} \oplus \mathbf{k}$ and the corresponding action for the Manin triple based on $\mathbf{k} \oplus \mathfrak{h}$, with $\alpha = -\bar{\rho}$.

Let us start with a very simple example of a Manin triple: the two-dimensional abelian Lie algebra with hyperbolic inner product $\mathbb{R}^{1,1}$. Let $\{x_+\}$ be a basis of $\mathfrak{g}_+ \cong \mathbb{R}$, and let $x^+$ be the vector such that $(x_+, x^+) = 1$. We introduce a pair of free fermionic fields $\{a_+(z), a^+(z)\}$, such that

$$a_+(z) \cdot a^+(w) \sim \frac{1}{z-w},$$

and a pair of free bosonic fields $\{\gamma, \beta\}$ such that

$$\beta(z) \cdot \gamma(w) \sim \frac{1}{z-w}.$$

We may realize the Heisenberg algebra $\widehat{\mathbb{R}^{1,1}}$ in a Fock space associated to the fields $\gamma(z)$ and $\beta(z)$ by means of currents $J_+ = \beta$ and $J^+ = \partial\gamma$. In this way, we obtain a free-field realization of the $N = 2$ superconformal algebra with central charge $d = 1$:

$$\begin{aligned}
\mathsf{G}^+ &= a^+\beta, \\
\mathsf{G}^- &= a_+\partial\gamma, \\
\mathsf{J} &= a^+a_+, \\
\mathsf{T} &= \partial\gamma\beta + \tfrac{1}{2}(\partial a^+ a_+ - a^+\partial a_+).
\end{aligned}$$

The associated BRS operator $\mathsf{Q} = \mathrm{Res}(a_+\beta)$ has one-dimensional cohomology, spanned by the vacuum $|\ \rangle$. (Physicists call the Fock space associated to the fields $\{\gamma, \beta, a^+, a_+\}$ a Kugo-Ojima quartet, while mathematicians call it a Koszul complex.)

Next, we turn to the the Manin triple $\mathfrak{g} = \mathbf{sl}(2) \oplus \mathbf{so}(2)$ associated to the Borel subalgebra $\mathbf{b}$ of $\mathbf{sl}(2)$. The Lie algebra $\mathbf{sl}(2)$ has basis

$$J_+ = \begin{pmatrix} 0 & 1 \\ 0 & 0 \end{pmatrix}, \quad J_0 = \begin{pmatrix} 1 & 0 \\ 0 & -1 \end{pmatrix}, \quad J_- = \begin{pmatrix} 0 & 0 \\ 1 & 0 \end{pmatrix}.$$

Denote by $\kappa$ the inner product $(J_+, J_-)$: since the dual Coxeter number of $\mathbf{sl}(2)$ is 2, this Manin triple corresponds to a representation of $\widehat{\mathbf{sl}(2)}$ at level $k = \kappa - 2$. Denote by $j$ the element of $\mathbf{so}(2)$ corresponding to the element $J_0 \in \mathbf{sl}(2)$. The subalgebra $\mathfrak{g}_+$ has basis

$$x_0 = \tfrac{1}{2}(J_0 + j), \quad x_- = J_-,$$

while the dual basis of the subalgebra $\mathfrak{g}_-$ is

$$x^0 = (2\kappa)^{-1}(J_0 - j), \quad x^- = \kappa^{-1}J_+.$$

The element $\rho$ equals $[x_0, x^0] + [x_-, x^-] = -\kappa^{-1}x_0 - x^0$, so that $\rho_+ = -\kappa^{-1}x_0$ and $\rho_- = -x^0$.

Let $J_\pm(z)$ and $J_0(z)$ be fields realizing a highest weight representation of $\widehat{\mathbf{sl}(2)}$ at level $k = \kappa - 2$:

$$\begin{aligned}
J_+(z) \cdot J_-(w) &\sim \frac{k}{(z-w)^2} + \frac{J_0(w)}{z-w}, \\
J_0(z) \cdot J_\pm(w) &\sim \pm\frac{2J_\pm(w)}{z-w}, \\
J_0(z) \cdot J_0(w) &\sim \frac{2k}{(z-w)^2}.
\end{aligned}$$

The field $j(z)$ is a free field with
$$j(z) \cdot j(w) \sim -\frac{2\kappa}{(z-w)^2}.$$

Introduce free fermionic fields $\{a_0(z), a^0(z), a_-(z), a^-(z)\}$, with
$$a_0(z) \cdot a^0(w) \sim a_-(z) \cdot a^-(w) \sim \frac{1}{z-w}.$$

Consider the following two-parameter family of actions of the $N=2$ superconformal algebra, with central charge $d = 3 - (1-st)/\kappa$:

$$\begin{aligned}
\mathsf{G}^+ &= a^+\beta + a^0\bigl(\tfrac{1}{2}(J_0+j) + a^-a_-\bigr) + a^- J_- + s\partial a^0, \\
\mathsf{G}^- &= a_+\partial\gamma + \kappa^{-1}a_0\bigl(\tfrac{1}{2}(J_0-j) + a^-a_-\bigr) + \kappa^{-1}a_- J_+ + t\partial a_0, \\
\mathsf{J} &= a^+a_+ + a^0 a_0 + a^- a_- - \kappa^{-1}(1+s)[\mathsf{Q},a_0] - (1-t)[\mathsf{Q},a^0]), \\
\mathsf{T} &= \beta\partial\gamma + (2\kappa)^{-1}\bigl(\tfrac{1}{2}(J_0)^2 + J_+J_- + J_-J_+ - \tfrac{1}{2}j^2 + s[\mathsf{Q},\partial a_0] + \kappa t[\mathsf{Q},\partial a^0]\bigr), \\
&\quad + \tfrac{1}{2}\bigl(\partial a^0 a_0 + \partial a^0 a_0 + \partial a^- a_- - a^0\partial a_0 - a^0\partial a_0 - a^-\partial a_-\bigr).
\end{aligned}$$

Here we have parametrized the element $\alpha \in \mathfrak{g}_0$ as $\alpha = sx_0 + \kappa t x^0$, the BRS operator $\mathsf{Q} = \operatorname{Res}(\mathsf{G}^+)$ is the zero-mode of the field $\mathsf{G}^+$, and

$$\begin{aligned}[]
[\mathsf{Q},a_0] &= \tfrac{1}{2}(J_0+j) + a^-a_-, \\
\kappa[\mathsf{Q},a^0] &= \tfrac{1}{2}(J_0-j) + a^-a_-.
\end{aligned}$$

For $s=t=0$, this is the tensor product of the actions of the $N=2$ superconformal algebra associated to the abelian Lie algebra $\mathbb{R}^{1,1}$ and the Kazama-Suzuki model at level $k$, associated to the Manin triple $\mathbf{sl}(2) \oplus \mathbf{so}(2)$. We will show that at $s=1$ and $t=-1$, this representation of the $N=2$ superconformal algebra is equivalent to that associated to the Manin triple $\mathbf{sl}(2) \oplus \mathbf{sl}(2)$: according to Karabali and Schnitzer [13] and Schnitzer [19], this describes the $\mathrm{SL}(2)/\mathrm{SL}(2)$ model at level $k$.

Consider the effect of twisting this action by the operator $\mathsf{R} = (2\kappa)^{-1}a_+a_-$. The conditions of Proposition 6.1 are satisfied, and we obtain an action of the $N=1\tfrac{1}{2}$ superconformal algebra where the fields $\mathsf{G}^+$, $\mathsf{J}$ and $\mathsf{T}$ are unchanged by the twist, while

$$\begin{aligned}
\tau_{\mathsf{R}}\mathsf{G}^- &= (2\kappa)^{-1}\bigl(a_+(J_- + 2\kappa\partial\gamma) + a_0(J_0 - j + 2a^-a_-) + a_-(2J_+ - \beta) \\
&\quad + a^0 a_+ a_- + 2\partial a_0\bigr), \\
\Phi &= -\frac{1}{3\kappa^2}a_+ a_0 a_-.
\end{aligned}$$

Of course, $\mathsf{F} = [\mathsf{Q},\Phi]$: we will not need its explicit formula.

We now introduce the field
$$\mathsf{A} = \gamma(J_+ - 2a^- a_0 + a^0 a_+).$$

The following operator products are easily demonstrated:

$$\mathsf{T}(z) \cdot \mathsf{A}(w) \sim \frac{\mathsf{A}(w)}{(z-w)^2} + \frac{\partial \mathsf{A}(w)}{z-w} - (s+t)\frac{\partial(\gamma(w)(J_+(w) - 2a^-(w)a_0(w)))}{(z-w)^2},$$

$$\mathsf{J}(z) \cdot \mathsf{A}(w) \sim (t-s-2)\frac{\gamma(w)(J_+(w) - 2a^-(w)a_0(w))}{z-w}.$$

In order for conjugation by the zero-mode of the field $\mathsf{A}$ to be physically meaningful, the field $\mathsf{A}$ should have conformal dimension 1 and charge 0. For this reason, we set $s = -1$ and $t = 1$ for the remainder of this section. For these values of the parameters, the fields $\mathsf{J}$ and $\mathsf{T}$ are given by the formulas

$$\mathsf{J} = a^+ a_+ + a^0 a_0 + a^- a_-,$$
$$\mathsf{T} = \beta \partial \gamma + (4\kappa)^{-1}\Big((J_0)^2 + 2J_+ J_- + 2J_- J_+ - j^2 - 2\partial j\Big),$$
$$+ \tfrac{1}{2}\Big(\partial a^0 a_0 + \partial a^0 a_0 + \partial a^- a_- - a^0 \partial a_0 - a^0 \partial a_0 - a^- \partial a_-\Big)$$

The Wakimoto representation of the affine Kac-Moody algebra at level $-k-4$ is realized in the Fock space of the free bosonic field $j(z)$ and free bosonic fields $\beta(z)$ and $\gamma(z)$, with currents

$$K_+ = \beta$$
$$K_0 = -2\gamma\beta + j$$
$$K_- = -\gamma^2 \beta + \gamma j - (k+4)\partial\gamma.$$

(The highest weight $\Lambda$ of the Wakimoto representation equals the vacuum expectation value of the zero-mode of $j$.)

**Theorem 7.1.**

$$e^{-\operatorname{ad}(\mathsf{A})}\mathsf{G}^+ = a^+(J_+ + K_+) + a^0\Big(\tfrac{1}{2}(J_0 + K_0) - a^+ a_+ + a^- a_-\Big) + a^-(J_- + K_-)$$
$$- 2a^+ a^- a_0,$$
$$e^{-\operatorname{ad}(\mathsf{A})}\mathsf{T}_\mathsf{R}\mathsf{G}^- = (2\kappa)^{-1}\Big(a_+(J_- - K_-) + a_0(J_0 - K_0) + a_-(J_+ - K_+)\Big),$$

*while the fields* $\mathsf{J}$, $\mathsf{T}$ *and* $\Phi$ *are left invariant by the automorphism* $e^{-\operatorname{ad}(\mathsf{A})}$.

*Proof.* Denote the field $J_+ - 2a^- a_0 + a^0 a_+$ by $I_+$. The calculation of $e^{-\operatorname{ad}(\mathsf{A})}\mathsf{G}^+$ follows from the formulas

$$\operatorname{ad}(\mathsf{A})\mathsf{G}^+ = -a^+ I_+ - a^0 \gamma(I_+ - \beta) + a^-(-\gamma j + (k+4)\partial\gamma) + 2\partial a^- \gamma - \partial a^0,$$
$$\tfrac{1}{2}\operatorname{ad}(\mathsf{A})^2 \mathsf{G}^+ = -a^0 \gamma I_+ + a^- \gamma^2(I_+ - \beta) + 2\partial a^- \gamma,$$
$$\tfrac{1}{6}\operatorname{ad}(\mathsf{A})^3 \mathsf{G}^+ = a^- \gamma^2 I_+, \quad \operatorname{ad}(\mathsf{A})^4 \mathsf{G}^+ = 0.$$

The calculation of $e^{-\operatorname{ad}(\mathsf{A})}\tau_{\mathsf{R}}\mathsf{G}^-$ follows from the formulas

$$\operatorname{ad}(\mathsf{A})\tau_{\mathsf{R}}\mathsf{G}^- = (2\kappa)^{-1}\Big(a_+(\gamma j + k\partial\gamma) + 2a_0\gamma(I_+ - \beta) + a_-I_+ - 2\partial a_+\gamma + 2\partial a_0\Big),$$

$$\tfrac{1}{2}\operatorname{ad}(\mathsf{A})^2\tau_{\mathsf{R}}\mathsf{G}^- = (2\kappa)^{-1}\Big(a_+\gamma^2(\beta - I_+) + 2a_0\gamma I_+ - 2\partial a_+\gamma\Big),$$

$$\tfrac{1}{6}\operatorname{ad}(\mathsf{A})^3\tau_{\mathsf{R}}\mathsf{G}^- = -(2\kappa)^{-1}a_+\gamma^2 I_+, \quad \operatorname{ad}(\mathsf{A})^4\tau_{\mathsf{R}}\mathsf{G}^- = 0.$$

The fields $\mathsf{J}$ and $\mathsf{T}$ are invariant under $\operatorname{ad}(\mathsf{A})$, since we have chosen $s = -1$ and $t = 1$, while the invariance of the field $\Phi$ is easily seen to satisfy $\operatorname{ad}(\mathsf{A})\Phi = 0$. $\square$

As a last transformation, let us apply the twist $\tau_{-\mathsf{R}}$ to the the above action of the $N = 1\tfrac{1}{2}$ superconformal algebra. Once more, the conditions of Proposition 6.1 are satisfied. The resulting action actually has $N = 2$ superconformal symmetry, with action

$$\mathbb{G}^+ = e^{-\operatorname{ad}(\mathsf{A})}\mathsf{G}^+ = a^+(J_+ + K_+) + a^0\big(\tfrac{1}{2}(J_0 + K_0) - a^+a_+ + a^-a_-\big) + a^-(J_- + K_-)$$
$$\phantom{\mathbb{G}^+ =}- 2a^+a^-a_0,$$

$$\mathbb{G}^- = \tau_{-\mathsf{R}}e^{\operatorname{ad}(-\mathsf{A})}\tau_{\mathsf{R}}\mathsf{G}^- = \kappa^{-1}\Big(-a_+K_- + a_0(\tfrac{1}{2}(J_0 - K_0) + a^+a_+ + a^-a_-) + a_-J_+ + \partial a_0)\Big),$$

$$\mathbb{T} = \mathsf{T} + (2\kappa)^{-1}[\mathbb{Q}, \partial a_0],$$

$$\mathbb{J} = \mathsf{J} + \kappa^{-1}[\mathbb{Q}, a_0],$$

where $\mathbb{Q}$ is the zero mode of $\mathbb{G}^+$, and

$$[\mathbb{Q}, a_0] = \tfrac{1}{2}(J_0 - K_0) - a^+a_+ + a^-a_-.$$

This is the $N = 2$ action associated to the Manin triple $\mathfrak{g} = \mathbf{sl}(2) \oplus \mathbf{sl}(2)$: $\mathfrak{g}_+$ has basis

$$x_+ = J_+ + K_+, \quad x_0 = \tfrac{1}{2}(J_0 + K_0), \quad x_1 = J_- + K_-,$$

$\mathfrak{g}_-$ has dual basis

$$x^+ = -\kappa^{-1}K_-, \quad x^0 = (2\kappa)^{-1}(J_0 - K_0), \quad x^- = \kappa^{-1}J_+,$$

the subspace $\mathfrak{g}_0$ is spanned by $x_0$, and the element $\alpha$ is taken to equal $\kappa^{-1}x_0$.

## Appendix. The operator product expansion and vertex algebras

In this paper, we have made free use of the operator product expansion of conformal field theory, as axiomatized in the notion of a vertex algebra. A vertex algebra is a $\mathbb{Z}/2$-graded vector space $V$, together with an even map $\partial : V \to V$ (which may be interpreted as differentiation), and a sequence of bilinear products

$$[AB]_n : V \otimes V \to V, \quad n \in \mathbb{Z},$$

such that for given $A$ and $B$, the product $[AB]_n$ vanishes for $n$ sufficiently large. These products are the Laurent coefficients of the operator product of the two fields in two-dimensional conformal field theory (Belavin-Polyakov-Zamolodchikov [3]):

$$A(z) \cdot B(w) = \sum_{n=-\infty}^{\infty} \frac{[AB]_n(w)}{(z-w)^n}.$$

For this reason, elements of a vertex algebra are sometimes referred to as fields. The product $[AB]_0$ is called the normal product of $A$ and $B$, and should be thought of as a renormalized product on $V$.

These products satisfy axioms, first written down by Borcherds [4], which are reminiscent of those of those of a Lie algebra. (We have reindexed his products in order to agree with the conventions of physicists: thus, we write $[AB]_n$ where Borcherds writes $A_{n-1}B$.)

**(Jacobi)**

$$[[AB]_m C]_n = \sum_{i=0}^{\infty} (-1)^i \binom{m-1}{i} \Big([A[BC]_{n+i}]_{m-i} + (-1)^{m+|A||B|}[B[AC]_{i+1}]_{m+n-i-1}\Big)$$

Here $\binom{a}{i} = a(a-1)\dots(a-i+1)/i!$.

**(Commutativity)**

$$[BA]_n = (-1)^{n+|A||B|} \sum_{i=0}^{\infty} \frac{(-1)^i}{i!} \partial^i [AB]_{n+i}$$

**(Identity)** There is an even element 1 such that $\partial 1 = 0$ and for all $A \in V$,

$$[1A]_n = \begin{cases} A, & n = 0, \\ 0, & n \neq 0. \end{cases}$$

For $m = n = 0$, the Jacobi rule simply says that

$$[[AB]_0 C]_n = \sum_{i \leq 0} [A[BC]_{-i}]_i + \sum_{i > 0} (-1)^{|A||B|} [B[AC]_i]_{-i}.$$

Subtracting the corresponding formula for $(-1)^{|A||B|}[[BA]_0 C]_0$, we see that

$$[[AB]_0 C]_n - (-1)^{|A||B|}[[BA]_0 C]_0 = [A[BC]_0]_0 - (-1)^{|A||B|}[B[AC]_0]_0.$$

The above axioms imply the following properties of the derivation $\partial$.

**Proposition A.1.** $[(\partial A)B]_n = (1-n)[AB]_{n-1}$ and $\partial[AB]_n = [(\partial A)B]_n + [A(\partial B)]_n$

*Proof.* To calculate $[(\partial A)B]_n$, we use the Jacobi rule, and the formula $\partial A = [A1]_{-1}$:

$$[(\partial A)B]_n = [[A1]_{-1}B]_n = \sum_{i=0}^{\infty} (-1)^i \binom{-2}{i} \Big([A[1B]_{n+i}]_{-i-1} - [1[AB]_{i+1}]_{n-i-2}\Big).$$

Note that $(-1)^i \binom{-2}{i} = i+1$. If $n < 1$, only the first term on the left-hand side contributes, with $i = -n$. On the other hand, if $n > 1$, only the second term contributes, with $i = n - 2$.

To calculate $[A(\partial B)]_n$, we apply the commutativity axiom:

$$[A(\partial B)]_n = \sum_{i=0}^{\infty} \frac{(-1)^{n+i+|A||B|}}{i!} \partial^i[(\partial B)A]_{n+i}$$

$$= \sum_{i=0}^{\infty} \frac{(-1)^{n+i+|A||B|}}{i!}(1 - n - i)\partial^i[BA]_{n+i-1}$$

$$= (n-1)\sum_{i=0}^{\infty} \frac{(-1)^{(n-1)+i+|A||B|}}{i!} \partial^i[BA]_{n+i-1} + \sum_{i=0}^{\infty} \frac{(-1)^{n+i+|A||B|}}{i!} \partial^i[BA]_{n+i}$$

$$= (n-1)[AB]_{n-1} + \partial[AB]_n. \quad \square$$

In this paper, we will only use the following consequence of the Jacobi rule, discovered by Sevrin et al. [21]. This formula shows the extent to which the map $B \mapsto [AB]_m$, $m > 0$, behaves like a derivation.

**Proposition A.2.** *If $m > 0$,*

$$[A[BC]_n]_m = \sum_{i=0}^{m-1} \binom{m-1}{i}[[AB]_{m-i}C]_{n+i} + (-1)^{|A||B|}[B[AC]_m]_n$$

$$= \sum_{i=0}^{\infty} \frac{(-1)^i}{i!}[[(\partial^i A)B]_m C]_{n+i} + (-1)^{|A||B|}[B[AC]_m]_n.$$

*Proof.* If $m > 0$, the Jacobi rule may be rewritten

$$[[AB]_m C]_n = \sum_{i=0}^{m-1}(-1)^i\binom{m-1}{i}\Big([A[BC]_{n+i}]_{m-i} - (-1)^{|A||B|}[B[AC]_{m-i}]_{n+i}\Big).$$

It follows that

$$\sum_{j=0}^{m-1}\binom{m-1}{j}[[AB]_{m-j}C]_{n+j} = \sum_{j=0}^{m-1}\sum_{i=0}^{m-j-1}(-1)^i\binom{m-1}{j}\binom{m-j-1}{i}$$

$$\Big([A[BC]_{n+i+j}]_{m-i-j} - (-1)^{|A||B|}[B[AC]_{m-i-j}]_{n+i+j}\Big).$$

The proposition now follows from the combinatorial identity

$$\sum_{\{0 \leq i,j \leq m-1 | i+j=k\}} (-1)^i\binom{m-1}{j}\binom{m-j-1}{i} = \begin{cases} 1, & k = 0, \\ 0, & k > 0, \end{cases}$$

which is a trivial consequence of the identity $(x + (1-x))^{m-1} = 1$. $\square$

The singular part of the operator product expansion is denoted

$$A(z) \cdot B(w) \sim \sum_{n>0} \frac{[AB]_n(w)}{(z-w)^n}.$$

We will call the algebraic structure similar to a vertex algebra, but with products $[AB]_n$ only for $n > 0$, satisfying the commutativity and identity axioms as well as the formula of Propositions A.2 and A.1, a chiral algebra. The reader is warned that this is not a standard piece of terminology, but it will be useful to have a word for this notion.

A stress-energy tensor in a vertex algebra is an even element $\mathsf{T}$ such that for all $A \in V$,
$$[\mathsf{T}A]_1 = \partial A,$$
and for some scalar $c$ (the central charge of $\mathsf{T}$),
$$\mathsf{T}(z) \cdot \mathsf{T}(w) \sim \frac{\frac{1}{2}c \cdot 1(w)}{(z-w)^4} + \frac{2\mathsf{T}(w)}{(z-w)^2} + \frac{\partial \mathsf{T}(w)}{z-w}.$$
It is usual to denote the scalar multiples $c \cdot 1(w)$ of the identity 1 simply by $c$. An element $A$ of $V$ has conformal dimension $a \in \mathbb{R}$ if $[\mathsf{T}A]_2 = aA$. For example, the conformal dimension of 1 is zero, while the conformal dimension of $\mathsf{T}$ is 2.

**Lemma A.3.** *The conformal dimension of $[AB]_0$ is the sum of the conformal dimensions of $A$ and $B$.*

*Proof.* By the Leibniz rule, we see that
$$[\mathsf{T}[AB]_0]_2 = [[\mathsf{T}A]_2 B]_0 + [[\mathsf{T}A]_1 B]_1 + [A[\mathsf{T}B]_2]_0 = (a+b)[AB]_0 + [(\partial A)B]_1.$$
The second term vanishes by Proposition A.1. □

To a chiral algebra $V$ (and in particular to a vertex algebra) is associated a Lie algebra
$$L(V) = V[z, z^{-1}]/\{(\partial A)z^n + nAz^{n-1}\},$$
with bracket
$$[Az^n, Bz^m] = \sum_{k \geq 0} \binom{n}{k}[AB]_{k+1} z^{n+m-k}.$$
If $A$ has conformal dimension $a$, one denotes the element $Az^n$ of the Lie algebra $L(V)$ by $A_{n-a+1}$. It follows from the properties of the identity that $1_n = 0$ for $n \neq 0$, and that $1_0$ lies in the centre of $L(V)$. The modes of the stress-energy tensor satisfy the relations of the Virasoro algebra,
$$[\mathsf{T}_n, \mathsf{T}_m] = (n-m)\mathsf{T}_{n+m} + \frac{c}{12}n(n^2-1)1_{n+m}.$$

By a representation of a chiral algebra, we will mean a representation of its associated Lie algebra $L(V)$. We only consider representations in the category $\mathcal{O}$, the definition of which is analogous to that of the category $\mathcal{O}$ for an affine Kac-Moody algebra or the Virasoro algebra.

If $K$ is a field of conformal dimension 1, then given fields $A$ and $B = [KA]_1$, we have the relation $B_n = [K_0, A_n]$. For this reason, we denote the graded commutator with the zero mode $K_0$ by $\mathrm{ad}(K)$.

An element $A$ is a primary field of conformal dimension $a$ if

$$\mathsf{T}(z) \cdot A(w) \sim \frac{aA(w)}{(z-w)^2} + \frac{\partial A(w)}{z-w}.$$

The modes of a primary field $A$ in $L(V)$ satisfy the following commutation relations with the Virasoro algebra spanned by $\mathsf{T}_n$:

$$[\mathsf{T}_n, A_m] = ((a-1)n - m)A_{n+m}.$$

We can now give the proof of Proposition 1.2.

*Proof.* The proof is a straighforward, though lengthy, exercise in the application of Proposition A.2. We will use the formulas

$$\mathsf{T} = [\mathsf{G}^+\mathsf{G}^-]_1 - \tfrac{1}{2}\partial\mathsf{J} = [\mathsf{G}^-\mathsf{G}^+]_1 + \tfrac{1}{2}\partial\mathsf{J} = \tfrac{1}{2}[\mathsf{G}^+\mathsf{G}^-]_1 + \tfrac{1}{2}[\mathsf{G}^-\mathsf{G}^+]_1,$$
$$\mathsf{J} = [\mathsf{G}^+\mathsf{G}^-]_2 = -[\mathsf{G}^-\mathsf{G}^+]_2,$$
$$d = [\mathsf{G}^+\mathsf{G}^-]_3 = [\mathsf{G}^-\mathsf{G}^+]_3.$$

(1) We start by calculating $\mathsf{J}(z) \cdot \mathsf{J}(w)$:

$$[\mathsf{J}\mathsf{J}]_n = [\mathsf{J}[\mathsf{G}^+\mathsf{G}^-]_2]_n = [[\mathsf{J}\mathsf{G}^+]_1\mathsf{G}^-]_{n+1} + [\mathsf{G}^+[\mathsf{J}\mathsf{G}^-]_n]_2$$

$$= \begin{cases} [[\mathsf{J}\mathsf{G}^+]_1\mathsf{G}^-]_2 + [\mathsf{G}^+[\mathsf{J}\mathsf{G}^-]_1]_2 = 0, & n = 1, \\ [[\mathsf{J}\mathsf{G}^+]_1\mathsf{G}^-]_3 = d, & n = 2, \\ [[\mathsf{J}\mathsf{G}^+]_1\mathsf{G}^-]_{n+1} = 0, & n > 2. \end{cases}$$

(2) Next we calculate $\mathsf{J}(z) \cdot \mathsf{T}(w)$:

$$[\mathsf{J}\mathsf{T}]_n = \tfrac{1}{2}([\mathsf{J}[\mathsf{G}^+\mathsf{G}^-]_1]_n + [\mathsf{J}[\mathsf{G}^-\mathsf{G}^+]_1]_n)$$
$$= \tfrac{1}{2}([[\mathsf{J}\mathsf{G}^+]_1\mathsf{G}^-]_n + [[\mathsf{J}\mathsf{G}^-]_1\mathsf{G}^+]_n) + \tfrac{1}{2}([\mathsf{G}^+[\mathsf{J}\mathsf{G}^-]_n]_1 + [\mathsf{G}^-[\mathsf{J}\mathsf{G}^+]_n]_1)$$

$$= \begin{cases} \tfrac{1}{2}([[\mathsf{J}\mathsf{G}^+]_1\mathsf{G}^-]_1 + [[\mathsf{J}\mathsf{G}^-]_1\mathsf{G}^+]_1 + [\mathsf{G}^+[\mathsf{J}\mathsf{G}^-]_1]_1 + [\mathsf{G}^-[\mathsf{J}\mathsf{G}^+]_1]_1) = 0, & n = 1, \\ \tfrac{1}{2}([[\mathsf{J}\mathsf{G}^+]_1\mathsf{G}^-]_2 + [[\mathsf{J}\mathsf{G}^-]_1\mathsf{G}^+]_2) = \mathsf{J}, & n = 2, \\ \tfrac{1}{2}([[\mathsf{J}\mathsf{G}^+]_1\mathsf{G}^-]_3 + [[\mathsf{J}\mathsf{G}^-]_1\mathsf{G}^+]_3) = d - d = 0, & n = 3, \\ \tfrac{1}{2}([[\mathsf{J}\mathsf{G}^+]_1\mathsf{G}^-]_n + [[\mathsf{J}\mathsf{G}^-]_1\mathsf{G}^+]_n) = 0, & n > 3. \end{cases}$$

(3) Next we show that $\mathsf{G}^+(z) \cdot \mathsf{F}(w) = 0$. Observe that

$$[\mathsf{G}^+\mathsf{F}]_n = [\mathsf{G}^+[\mathsf{G}^+\Phi]_1]_n = -[\mathsf{G}^+[\mathsf{G}^+\Phi]_n]_1.$$

The right-hand side vanishes for $n > 1$, while it equals the negative of the left-hand side for $n = 1$.

(4) We next calculate $\mathsf{G}^-(z)\cdot\mathsf{T}(w)$. We will use the fact that

$$[\mathsf{G}^\pm(\partial\mathsf{J})]_n = \partial[\mathsf{G}^\pm\mathsf{J}]_n + [(\partial\mathsf{G}^\pm)\mathsf{J}]_n = \begin{cases} \mp\partial\mathsf{G}^\pm, & n=1, \\ \pm\mathsf{G}^\pm, & n=2, \\ 0, & n>2. \end{cases}$$

Also, observe that

$$[\mathsf{G}^-[\mathsf{G}^-\mathsf{G}^+]_1]_n = [[\mathsf{G}^-\mathsf{G}^-]_1\mathsf{G}^+]_n - [\mathsf{G}^-[\mathsf{G}^-\mathsf{G}^+]_n]_1.$$

For $n=1$, this shows that $[\mathsf{G}^-[\mathsf{G}^-\mathsf{G}^+]_1]_1 = \frac{1}{2}[\mathsf{F}\mathsf{G}^+]_1 = 0$, and thus that

$$[\mathsf{G}^-\mathsf{T}]_n = [\mathsf{G}^-[\mathsf{G}^-\mathsf{G}^+]_1]_n + \tfrac{1}{2}[\mathsf{G}^-(\partial\mathsf{J})]_n = \tfrac{1}{2}\partial\mathsf{G}^-.$$

For $n=2$, we see that

$$[\mathsf{G}^-\mathsf{T}]_2 = [\mathsf{G}^-[\mathsf{G}^-\mathsf{G}^+]_1]_2 + \tfrac{1}{2}[\mathsf{G}^-(\partial\mathsf{J})]_2 = -[\mathsf{G}^-[\mathsf{G}^-\mathsf{G}^+]_2]_1 - \tfrac{1}{2}\mathsf{G}^- = -\tfrac{3}{2}\mathsf{G}^-.$$

Similar reasoning shows that $[\mathsf{G}^-\mathsf{T}]_n = 0$ for $n>2$.

The calculation of $\mathsf{G}^+(z)\cdot\mathsf{T}(w)$ is similar, except that we use the formula $\mathsf{T} = [\mathsf{G}^+\mathsf{G}^-]_1 - \tfrac{1}{2}\partial\mathsf{J}$.

(5) Next, we calculate $\mathsf{T}(z)\cdot\mathsf{T}(w)$:

$$\begin{aligned}[\mathsf{TT}]_n &= \tfrac{1}{2}([\mathsf{T}[\mathsf{G}^+\mathsf{G}^-]_1]_n + [\mathsf{T}[\mathsf{G}^-\mathsf{G}^+]_1]_n) \\ &= \tfrac{1}{2}((n-1)[[\mathsf{TG}^+]_2\mathsf{G}^-]_{n-1} + [[\mathsf{TG}^+]_1\mathsf{G}^-]_n \\ &\quad + (n-1)[[\mathsf{TG}^-]_2\mathsf{G}^+]_{n-1} + [[\mathsf{TG}^-]_1\mathsf{G}^+]_n) \\ &\quad + \tfrac{1}{2}([\mathsf{G}^+[\mathsf{TG}^-]_n]_1 + [\mathsf{G}^-[\mathsf{TG}^+]_n]_1).\end{aligned}$$

For $n=1$, this equals

$$\tfrac{1}{2}\big([(\partial\mathsf{G}^+)\mathsf{G}^-]_1 + [(\partial\mathsf{G}^-)\mathsf{G}^+]_1 + [\mathsf{G}^+(\partial\mathsf{G}^-)]_1 + [\mathsf{G}^-(\partial\mathsf{G}^+)]_1\big) \\ = \tfrac{1}{2}(\partial[\mathsf{G}^+\mathsf{G}^-]_1 + \partial[\mathsf{G}^-\mathsf{G}^+]_1) = \partial\mathsf{T}.$$

For $n=2$, we obtain

$$\tfrac{1}{2}\big([[\mathsf{TG}^+]_2\mathsf{G}^-]_1 + [[\mathsf{TG}^-]_2\mathsf{G}^+]_1 + [[\mathsf{TG}^+]_1\mathsf{G}^-]_2 + [[\mathsf{TG}^-]_1\mathsf{G}^+]_2 \\ + [\mathsf{G}^+[\mathsf{TG}^-]_2]_1 + [\mathsf{G}^-[\mathsf{TG}^+]_2]_1\big) = 2\mathsf{T}.$$

For $n=3$, we obtain

$$\tfrac{1}{2}\big(2[[\mathsf{TG}^+]_2\mathsf{G}^-]_2 + 2[[\mathsf{TG}^-]_2\mathsf{G}^+]_2 + [[\mathsf{TG}^+]_1\mathsf{G}^-]_3 + [[\mathsf{TG}^-]_1\mathsf{G}^+]_3\big) = 0.$$

Finally, for $n=4$ we obtain

$$\tfrac{1}{2}\big(3[[\mathsf{TG}^+]_2\mathsf{G}^-]_3 + 3[[\mathsf{TG}^-]_2\mathsf{G}^+]_2 + [[\mathsf{TG}^+]_1\mathsf{G}^-]_4 + [[\mathsf{TG}^-]_1\mathsf{G}^+]_4\big) = \tfrac{3}{2}d.$$

(6) From the vanishing of $\mathsf{G}^-(z)\cdot\Phi(w)$, it follows that $[\Phi\mathsf{F}]_n = -\tfrac{1}{2}[\Phi[\mathsf{G}^-\mathsf{G}^-]_1]_n = 0$ for all $n>0$. From this, we see that $[\mathsf{FF}]_n = [\mathsf{G}^+[\Phi\mathsf{F}]_n]_1 = 0$.

(7) We now calculate $\mathsf{G}^-(z)\cdot\mathsf{F}(w)$. Using the fact $[\mathsf{J}\Phi]_1 = -3\Phi$, we see that
$$3\Phi = [\Phi\mathsf{J}]_1 = -[\Phi[\mathsf{G}^-\mathsf{G}^+]_2]_1 = [\mathsf{G}^-[\Phi\mathsf{G}^+]_1]_2 = [\mathsf{G}^-\mathsf{F}]_2.$$

Incidentally, this shows that
$$[\Phi\Phi]_n = \tfrac{1}{3}[\Phi[\mathsf{G}^-\mathsf{F}]_2]_n = -\tfrac{1}{3}[\mathsf{G}^-[\Phi\mathsf{F}]_n]_2 = 0.$$

A similar calculation shows that $[\mathsf{G}^-\mathsf{F}]_n = 0$ for $n > 2$. Since
$$[\mathsf{G}^-\mathsf{F}]_1 = -\tfrac{1}{2}[\mathsf{G}^-[\mathsf{G}^-\mathsf{G}^-]_1]_1 = -\tfrac{1}{2}[[\mathsf{G}^-\mathsf{G}^-]_1\mathsf{G}^-]_1 + \tfrac{1}{2}[\mathsf{G}^-[\mathsf{G}^-\mathsf{G}^-]_1]_1,$$

we see that $2[\mathsf{G}^-\mathsf{F}]_1 = [\mathsf{F}\mathsf{G}^-]_1$. But $[\mathsf{F}\mathsf{G}^-]_1 + [\mathsf{G}^-\mathsf{F}]_1 = \partial[\mathsf{G}^-\mathsf{F}]_2 = 3\partial\Phi$, and we conclude that $[\mathsf{G}^-\mathsf{F}]_1 = \partial\Phi$.

(8) We now calculate $\Phi(z)\cdot\mathsf{T}(w)$:
$$[\Phi\mathsf{T}]_n = \tfrac{1}{2}([\Phi[\mathsf{G}^+\mathsf{G}^-]_1]_n + [\Phi[\mathsf{G}^-\mathsf{G}^+]_1]_n) = \begin{cases} \tfrac{1}{2}([\mathsf{F}\mathsf{G}^-]_1 - [\mathsf{G}^-\mathsf{F}]_1) = \tfrac{1}{2}\partial\Phi, & n = 1, \\ \tfrac{1}{2}[\mathsf{F}\mathsf{G}^-]_2 = \tfrac{3}{2}\mathsf{F}, & n = 2, \\ \tfrac{1}{2}[\mathsf{F}\mathsf{G}^-]_n = 0, & n > 2. \end{cases}$$

The formula for $\mathsf{T}(z)\cdot\mathsf{F}(w)$ is easily obtained from this, since $\mathsf{F} = [\mathsf{G}^+\Phi]_1$. $\square$

Department of Mathematics, MIT, Cambridge MA 02139 USA
*E-mail address*: getzler@math.mit.edu